\documentclass[a4paper,11pt]{article}

\usepackage{jheppub} 
\usepackage[T1]{fontenc} 
\usepackage{amssymb}
\usepackage{amsmath}
\usepackage{amstext}
\usepackage{amsfonts}
\usepackage{comment}

\title{\boldmath Generalised Boundary Terms for Higher Derivative Theories of Gravity}

\author{Ali Teimouri$^{a}$,}
\author{Spyridon Talaganis$^{a}$,}
\author{James Edholm$^{a}$,}
\author{Anupam Mazumdar$^{a, b}$}

\affiliation[a]{Consortium for Fundamental Physics, \\ Lancaster University, \\Lancaster, LA1 4YB, United Kingdom.}
\affiliation[b]{Kapteyn Astronomical Institute, \\ University of Groningen, \\ 9700 AV Groningen, The Netherlands.}

\emailAdd{a.teimouri@lancaster.ac.uk}
\emailAdd{s.talaganis@lancaster.ac.uk}
\emailAdd{j.edholm@lancaster.ac.uk}
\emailAdd{a.mazumdar@lancaster.ac.uk}

\abstract{In this paper we wish to find the corresponding Gibbons-Hawking-York term for the most general quadratic in curvature gravity
by using Coframe slicing within the Arnowitt-Deser-Misner (ADM) decomposition of spacetime in four dimensions. In order to make sure
that the higher derivative gravity is {\it ghost and tachyon free} at a perturbative level, one requires infinite covariant derivatives, which yields a generalised
covariant infinite derivative theory of gravity. We will be exploring the boundary term for such a covariant infinite derivative theory of gravity.}

\begin{document} 

\maketitle
\flushbottom
\section{Introduction}\label{intro}

Einstein's General theory of Relativity (GR) has seen tremendous success
in matching its predictions with observations in the {\it weak field regime}
in the infrared (IR)~\cite{Will}, including the recent confirmation of the detection
of Gravitational Waves~\cite{Abbott:2016blz}.  However the same cannot be said for the
ultraviolet (UV), because 
at {\it small time scales} and {\it short distances}, the theory breaks down. At a classical
level, GR permits a solution with a blackhole singularity and a cosmological singularity~\cite{HawkingandEllis}.
Besides these classical singularities, the GR also begs for a quantum UV
completion. For instance, a pure quadratic gravity is 
renormalizable~\cite{stelle}, but being a higher derivative theory, the theory
suffers from {\it ghosts}, which can not be cured order by order.
In a classical context, higher derivatives also bring in an instability known as the
Ostrogradsky instability~\cite{ostro}.
In any case, gravity being a diffeomorphism invariant  theory, one would
naturally expect higher order {\it quantum} corrections which would 
allow all possible diffeomorphism invariant terms, such as covariant higher and {\it infinite derivative}
contributions in the Ricci scalar, Ricci tensor and Weyl, an explicit computation has been performed in~\cite{Biswas:2011ar}.

The covariant construction of {\it infinite derivative theories of gravity} (IDG), which is parity invariant and 
torsion-free, was first studied around the Minkowski background in ~\cite{Biswas:2011ar}, and then generalised to constant
curvature backgrounds, such as deSitter (dS) and Anti-deSitter (AdS) backgrounds in~\cite{Biswas:2016etb}.

Since these theories
contain infinite covariant derivatives, there is no highest momentum operator and as a result their perturbative stability
cannot be analysed via Ostrogradsky analysis. One has to understand their tree level propagator and study what are 
the {\it true} dynamical degrees of freedom which propagate in the space time. If one maintains that the dynamics of the original theory of gravity 
is governed by massless gravitons, then IDG must {\it not} contain any other degrees of freedom, other than spin-2 and 
spin-0 components of the massless graviton~\cite{Rivers,peter}. In order to prohibit no new states in the propagator, one has to 
make sure that there are no zeroes in the complex plane, such that both spin-2 and spin-0 components remain massless.
IDG will inevitably introduce new poles in the propagator. However, infinite poles corresponding to infinite covariant derivatives
can be summed up as the exponent of an {\it entire function}~\cite{Tomboulis,Siegel:2003vt,Biswas:2005qr,Biswas:2011ar,Biswas:2013kla,Biswas:2016etb}. By definition, 
such a modification will not incur any new poles in the propagator and also ensures that the modified propagator has a correct IR limit, where 
one recovers the predictions of {\it pure} Einstein-Hilbert 
action at large distances from the source and large time scales. 

This is indeed encouraging as there exists a {\it non-singular blackhole solution}, at least at the level of linearized equations of motion for such covariant
infinite derivative theory of gravity in a static~\cite{Biswas:2005qr,Biswas:2011ar,Edholm:2016hbt}, and in a time dependent background~\cite{Frolov}.  
In both cases the universal feature is that at short distances there exists no singularity and at large distances the theory behaves like in the case of GR.

These classical results transcend through the quantum domain. In the UV, such modifications soften the UV divergences by weakening the 
graviton propagator~\cite{Tomboulis,Modesto,Anselmi:2015dsa,Moffat-qg,Bravinsky}.
Since gravitational interactions contain only derivatives, therefore the vertex operators in such theories are exponentially enhanced,
which modifies the counting of superficial degree of divergence, and also gives rise to {\it nonlocal} interactions, determined by the new scale 
$M\leq M_p\sim 2.4\times 10^{18}$~GeV in four dimensions. Indeed, with a similar procedure one can now go to any arbitrary dimensions. As such, there
is no restriction on dimensionality from perturbative unitarity. For an exponential suppressed graviton propagator, explicit $2$-loop computations have been performed
for a scalar analogue of a graviton, and it was found that the $2$-loop computation yields a UV finite result~\cite{Talaganis:2014ida}, along with quantum scattering, where the 
scattering amplitude does not grow with external momenta~\cite{Talaganis:2016ovm}.

Note that such nonlocality is a common thread for many approaches to quantum gravity,
such as loop quantum gravity (LQG)~\cite{Ashtekar,Nicolai:2005mc}, causal dynamical approach~\cite{Sorkin}, and string 
theory (ST)~\cite{ST}. In LQG and causal
dynamical approaches the basic formulation is based on 
nonlocal objects, such as Wilson loops and fluxes coming from the gravitational
field.  ST  introduces nonlocal interactions,
where classically strings and branes interact over a region in space. In
string field theory (SFT), the appearance of noncommutative 
geometry~\cite{Witten:1985cc}, $p$-adic strings~\cite{Freund:1987kt}, zeta
strings~\cite{Dragovich:2007wb}, and strings quantized on a random lattice~\cite{Douglas:1989ve,Biswas:2004qu}
introduce nonlocality. For a review on SFT, see \cite{Siegel:1988yz}.
An important common feature in all these approaches is the presence of an
{\it  infinite series of higher-derivative} terms incorporating the 
nonlocality in the form of an {\it exponential kinetic} correction.

It is also worth mentioning that in recent years there has been a growing
interest in infinite-derivative gravitational theories in not only addressing
the Big Bang singularity problem~\cite{Biswas:2005qr,Biswas:2010zk,Biswas:2012bp,Craps:2014wga,Conroy:2016sac},
but also in finding cosmological inflation and UV completeness of Starobinsky
model of inflation~\cite{Biswas:2005qr,Chialva:2014rla,Biswas:2013dry}, and \cite{Koshelev:2016xqb}.
In such classes of theory, the gravitational entropy~\cite{Iyer:1994ys,Jacobson}, of a static and axisymmetric metric receives zero contribution from the infinite 
derivative sector of the action, when no additional scalar propagating modes are introduced. In this case, the gravitational entropy is strictly given by the 
area-law arising solely from the contribution of the Einstein-Hilbert action~\cite{Conroy:2015wfa}.

Irrespective of classical or quantum computations, one of the key features of a covariant action is to have a well-posed 
boundary condition. In particular, in the Euclidean path integral approach - requiring such an action to be stationary, one also requires all the boundary terms to 
disappear on any permitted variation. For instance, calculating the black 
hole entropy using the Euclidean semiclassical approach shows that the entire contribution comes from the boundary term~\cite{Brown:1995su,hawking1,Hawking:1995fd}. 
It is well known that the variation of the Einstein-Hilbert (EH) action leads
to a boundary term that depends
not just on the metric, but also on the derivatives of the metric. This is
due to
the fact that the action itself depends on the metric, along with terms that depend linearly on the second derivatives.
Normally, in Lagrangian field theory, such linear second
derivative terms can be introduced or eliminated, by adding an  appropriate
boundary term to the action. 
In gravity, the fact that the second derivatives arise linearly
and also the existence of total derivative indicates that the second derivatives
are redundant in the sense that they can be eliminated by  integrating
by parts, or by adding an appropriate boundary  term. Indeed,
writing a boundary term for a gravitational action schematically confines
the non-covariant terms to the boundary~\cite{york1}.  For the EH action this geometrically
transparent, boundary term is given by the Gibbons-Hawking-York (GHY) boundary term
\cite{Gibbons:1976ue}. Adding this boundary to the bulk action results in
an elimination of the total derivative, as seen for $f(R)$ gravity~\cite{Guarnizo:2010xr,Madsen:1989rz}.

In the Hamiltonian formalism, obtaining the boundary terms for a gravitational action is vital. This is due to the fact that the boundary term ensures that the path integral for 
quantum gravity admits correct answers. As a result, in the late 1950s the 3+1 decomposition received a great deal
of attention;
Richard Arnowitt, Stanley Deser and Charles W. Misner (ADM)  showed \cite{Arnowitt:1962hi}
that upon decomposing spacetime such that for the four dimensional Einstein equation
we have three-dimensional surfaces (later to be defined as hypersurfaces)
and one fixed time coordinate for each slices. We can therefore formulate and recast
the Einstein equations in terms of the Hamiltonian and hence achieve a better
insight into GR.
 
In the ADM decomposition, one foliates the arbitrary region $\mathcal{M}$
of the space-time manifold with a family of spacelike hypersurfaces $\Sigma_{t}$,
one for each instant in time. It has been shown by the authors of~\cite{Deruelle:2009zk}
that one can decompose a gravitational
action, using the ADM formalism and without necessarily moving into the Hamiltonian
regime, such that we obtain the total derivative of the gravitational action. Using this powerful technique, 
one can eliminate this total derivative 
term by modifying the GHY term appropriately.

The aim of this paper is to find the corresponding GHY boundary term for a covariant IDG.
We start by providing a 
warm up example of how to obtain a boundary term for an infinite derivative,
massless scalar field theory. We then in section~\ref{sec3}   briefly review the boundary term for EH term and introduce infinite
derivative gravity. We then set our preliminaries by discussing the time slicing in section~\ref{sec4}, and reviewing how one may obtain the boundary terms in the $3+1$ formalism 
in~\ref{sec5}. We finally turn our attention to our gravitational action and find the appropriate boundary terms for such a theory in section \ref{sec6}.

\section{Warm up exercise: Infinite derivative massless scalar field
theory}\label{sec2}

Let us consider the following action of a generic scalar field $\phi$ of mass dimension 2:
\begin{eqnarray}\label{eq1}
S_\phi= \int d^{4} x  \, \phi \Box^{n} \phi,
\end{eqnarray}
where $\Box=\eta^{\mu\nu}\nabla_\mu\nabla_\nu$, where $\eta_{\mu\nu}$ is the Minkowski metric~\footnote{The $\Box$ term comes with a scale $\Box/M^2$,
where $M$ is a new scale below $M \leq M_p=(16\pi G)^{-1/2}\sim 2.4\times 10^{18}$~GeV in $4$ dimensions.
The physical significance of $M$ could be any scale beyond $10^{-2}$~eV, which arises from constraints on 
studying the $1/r$-fall of the Newtonian potential~\cite{Edholm:2016hbt}.
In our notation, we  suppress the scale $M$  in order not
to clutter our formulae for the rest of this paper. However, for any physical comparison one has to bring in the scale $M$ along with
$M_p$.} and $n\in \mathbb{N}_{>0}$.
Generalising, we have that $\Box^{n}=\prod_{i=1}^{n} \eta^{\mu_{i}\nu_{i}}\nabla_{\mu_{i}}\nabla_{\nu_{i}}$. The aim is to find the total derivative term for the above action.
We may vary the scalar field $\phi$ as: $\phi \rightarrow\ \phi + \delta \phi$.
Then
the variation of the action is given by
\begin{eqnarray}\label{eq2}
\delta S_\phi&=&\int d^{4} x \, \big[\delta\phi \Box^{n} \phi+\phi \delta(\Box^{n}
\phi)\big]\,,\nonumber\\
&=&\int d^{4} x \, \big[\delta\phi \Box^{n} \phi+\phi\Box^{n}
 \delta\phi\big]\,,\nonumber\\
 &=&\int d^{4} x \, \big[(2\Box^{n}\phi)\delta\phi+X\big]\,,
\end{eqnarray}
where now $X$ are the $2n$ total derivatives:
\begin{eqnarray}\label{eq3}
X&=&\int d^{4} x \, \big[\nabla_\mu(\phi\nabla^\mu\Box^{n-1}\delta\phi)-\nabla^\mu(\nabla_\mu\phi\Box^{n-1}\delta\phi)\\\nonumber
&+&\nabla_\lambda(\Box\phi\nabla^\lambda\Box^{n-2}\delta\phi)-\nabla^\lambda(\nabla_\lambda\phi\Box^{n-2}\delta\phi)+\cdots
+\nabla_\sigma(\Box^{n-1}\phi\nabla^\sigma\delta\phi)-\nabla^\sigma(\nabla_\sigma\Box^{n-1}\phi\delta\phi)\big].
\end{eqnarray}
where ``$\cdots$'' in the above equation indicates the intermediate terms. 

Let us now consider a more general case
\begin{eqnarray}\label{eq4}S_{\phi}=\int d^4x \, \phi\mathcal{F}(\Box)\phi,\end{eqnarray}
where $\mathcal{F}(\Box)=\sum^{\infty}_{n=0}c_n\Box^n$, where the `$c_n$'s are dimensionless coefficients. In this case
the total derivatives are given by
\begin{eqnarray}\label{eq5}
X=\sum^{\infty}_{n=1}c_n\int d^4x\sum^{2n}_{j=1}(-1)^{j-1}\nabla_\mu(\nabla^{(j-1)}\phi\nabla^{(2n-j)}\delta\phi),
\end{eqnarray}
where the superscript $\nabla^{(j)}$ indicate the number of covariant derivatives
acting to the right. Therefore, one can always determine the total derivative
for any given action, and one can then preserve or eliminate these terms depending on the purpose of the study. 
In the following sections we wish to address how one can obtain the total
derivative for a given gravitational action.

\section{Introducing Infinite Derivative Gravity}\label{sec3}

The gravitational action is built up of two main components, the bulk part
and the boundary part. In the simplest and the most well known
case~\cite{Gibbons:1976ue}, for the Einstein-Hilbert (EH) action, the boundary
term are the ones known as Gibbons-Hawking-York (GHY) term. We can write the total EH action in terms of the bulk part and the boundary part simply as  
\begin{eqnarray}\label{eq6}
S_G&=&S_{EH}+S_{B}\nonumber \\
&=&\frac{1}{16\pi G}\int_{\mathcal{M}} d^{4}x \, \sqrt{-g}  \mathcal{R}+ \frac{1}{8\pi G}\oint_{\partial\mathcal{M}}
d^{3}y \, \varepsilon\left\vert h \right\vert^{1/2}K\,,  
\end{eqnarray}
where $\mathcal{R}$ is the Ricci-scalar, and  $K$ is the trace of the extrinsic curvature with $K_{ij}\equiv -\nabla_{i}n_{j}$, $\mathcal{M}$ indicates the 4-dimensional region and $\partial\mathcal{M}$ 
denotes the $3$-dimensional boundary region. $h$ is the determinant of the induced metric on the hypersurface $\partial\mathcal{M}$ and $\varepsilon=n^\mu n_\mu=\pm1$, 
where $\varepsilon$ is equal to $-1$ for a spacelike hypersurface,  and is equal to $+1$ for a timelike hypersurface when we take the metric signature is ``mostly plus''; 
i.e. $(-,+,+,+)$. A unit normal $n_{\mu}$ can be introduced only if the hypersurface is not null, and $n^\mu$ is the normal vector to the hypersurface.

Indeed, one 
can derive the boundary term simply by using the variational principle. In
this case the action is varied with respect to the 
metric, and it produces a total-divergent term, which can be eliminated by the
variation of $S_B$, \cite{Gibbons:1976ue}. Finding the boundary terms for 
any action is an indication that the variation principle for the given theory
is well posed.

As mentioned earlier on, despite the many successes that the EH action brought in understanding the universe in IR regime, the UV sector of gravity requires corrections 
to be well behaved. The most general {\it covariant} action of gravity, which is quadratic in curvature, can be written as~\cite{Biswas:2013kla},
\begin{eqnarray}\label{eq7}
S&=&S_{EH}+S_{UV}\nonumber\\
&=&\frac{1}{16\pi G}\int d^{4}x \, \sqrt{-g}\Big[\mathcal{R}+\alpha\big(\mathcal{R}\mathcal{F}_{1}(\Box)\mathcal{R}
+\mathcal{R}_{\mu\nu}\mathcal{F}_{2}(\Box)\mathcal{R}^{\mu\nu}\nonumber\\
&&+\mathcal{R}_{\mu\nu\rho\sigma}\mathcal{F}_{3}(\Box)\mathcal{R}^{\mu\nu\rho\sigma}\big)\Big],\quad
\text{with} \quad \mathcal{F}_{i}(\Box)=\sum^{\infty}_{n=0}f_{i_{n}}\Box^{n} \,,
\end{eqnarray}
where $\alpha$ is a constant with mass dimension $-2$ and the `$f_{i_{n}}$'s are dimensionless coefficients. 
For the full equations of motion of such an action, see~\cite{Biswas:2013cha}.

Around Minkowski spacetime, the {\it ghost free} condition gives a constraint on the form factors ${\cal F}_{i}(\Box)$'s~\cite{Biswas:2011ar,Biswas:2010zk,Biswas:2013cha}, 
\begin{equation}
2{\cal F}_1(\Box)+{\cal F}_2(\Box)+ 2{\cal F}_3 (\Box)=0\,.
\end{equation}
Around  Minkowski spacetime the Weyl contribution vanishes, since the last term in the action can be recast in terms of Weyl,  one can take
${\cal F}_3(\Box)=0$, in which case the condition for a  ghost-free graviton propagator leads to a particular choice of the form factor~\cite{Biswas:2011ar,Biswas:2005qr}:
\begin{equation}
 {\cal F}_3=0 \Rightarrow {\cal
F}_1(\Box)=-\frac{1}{2}{\cal F}_2(\Box) \Rightarrow {\cal F}_2(\Box)={a(\Box)-1\over 2\Box}\,,
\end{equation}
where $a(\Box)$ is an exponential of an {\it entire function}, which does not contain any zeroes. 
A particularly simple class which mimics the stringy gaussian nonlocalities is given by~\cite{Siegel:1988yz,Biswas:2005qr,Biswas:2011ar,Biswas:2010zk,Biswas:2013cha}
\begin{equation}
\label{choice-a}
a(\Box)=e^{-{\Box\over M^2}}\,,
\end{equation}
where $M$ is the scale of nonlocality. The aim of this paper is to seek the 
the boundary terms corresponding to $S_{UV}$, while retaining the Riemann term. A technically challenging question, but the answer will
have tremendous impact on various aspects of gauge theory and gravity which should be explored in future.

\section{Time Slicing}\label{sec4}

Any geometric spacetime can be recast in terms of time like spatial slices,
known as hypersurfaces.
How these slices are embedded in spacetime, determines the extrinsic curvature
of the slices. One of the motivations of time slicing is to evolve the 
equations of motion from a well-defined set of initial conditions set at a well-defined spacelike 
hypersurface, see~\cite{Smarr:1977uf,Baumgarte:2002jm}.

\subsection{ADM Decomposition}\label{sec4.1}

In order to define the decomposition, we first look at the foliation.
Suppose that the time orientable spacetime $\mathcal{M}$ is foliated by a
family of spacelike
hypersurfaces $\Sigma_{t}$, on which time is a fixed constant $t=x^{0}$.
We then define the induced metric on the hypersurface as $h_{ij}\equiv \left.g_{ij}\right |_{_t}$,
where the Latin indices run from $1$ to $3$~\footnote{ It should also be noted that Greek indices run from $0$ to $3$ and Latin indices run from $1$ to $3$, that is, only spatial coordinates are considered.}. The line element is then given
by the ADM (Arnowitt-Deser-Misner) decomposition~\cite{Relativiststoolkit}:
\begin{eqnarray} \label{eq8}
ds^{2}=-(N^{2}-\beta_{i}\beta^{i})dt^{2}+2\beta_{i}dx^{i}dt+h_{ij}dx^{i}dx^{j}
\end{eqnarray}
where $$N=\frac{1}{\sqrt{-g^{00}}}$$  is the ``lapse'' function, and $$\beta^{i}=-\frac{g^{0i}}{g^{00}}$$
is  the ``shift'' vector.
We may then define $n^\mu$, the vector normal to the hypersurface, as:
\begin{eqnarray}\label{eq9}
n_{i}=0, \quad n^{i}=-\frac{g^{0i}}{\sqrt{-g^{00}}},\quad
n_{0}=-\frac{1}{\sqrt{-g^{00}}},\quad\ n^0=\sqrt{-g^{00}}\,,
\end{eqnarray}
where for the ADM metric, given in  Eq.~(\ref{eq8}), $n^\mu$ takes the following
form:
\begin{eqnarray}\label{eq10}
n_{i}=0, \quad n^{i}=-\frac{\beta^{i}}{N},\quad
n_{0}=-N,\quad\ n^0=N^{-1}\,.
\end{eqnarray}
In the above line element Eq.~(\ref{eq8}), we also have $\sqrt{-g}=N\sqrt{h}$.  The induced metric of
the hypersurface can be related to the $4$ dimensional full metric via the completeness relation, where, for a spacelike hypersurface,
\begin{eqnarray}\label{eq11}
g^{\mu\nu}&=&h^{ij}e^{\mu}_{i}e^{\nu
}_{j}+\varepsilon n^{\mu}n^{\nu}\nonumber\\
&=& h^{\mu \nu}-n^{\mu}n^{\nu} \,,
\end{eqnarray}
where $\varepsilon=-1$ for a spacelike hypersurface, and $+1$ for a timelike hypersurface, and
\begin{equation}\label{eee}
e^\mu_i=\frac{\partial x^{\mu}}{\partial y^{i}}\,,
\end{equation}
are basis vectors on the hypersurface which allow us to define tangential tensors on the
  hypersurface\footnote{We can use $h^{\mu\nu}$ to project a tensor $A_{\mu\nu}$ onto the hypersurface: $A_{\mu\nu} e^\mu_i e^\nu_j = A_{ij}$ where $A_{ij}$ is the three-tensor associated with $A_{\mu\nu}$.}. We note $`x$'s are coordinates on region $\mathcal{M}$, while $`y$'s are coordinates associated with the hypersurface and we may also keep in mind that,
\begin{equation}\label{eq12}
h^{\mu\nu}=h^{ij}e^{\mu}_{i}e^{\nu
}_{j}\,,
\end{equation}
where $h^{ij}$ is the inverse of the induced metric $h_{ij}$ on the hypersurface, see for instance~\cite{Relativiststoolkit}.

The change of direction
of the normal $n$ as one moves on the hypersurface corresponds to the bending
of the hypersurface $\Sigma_{t}$  which is described by the extrinsic curvature.
The extrinsic curvature of spatial slices where time is constant is given by:
\begin{eqnarray}\label{eq13}
K_{ij}\equiv -\nabla_{i}n_{j}=\frac{1}{2N}\left(D_{i}\beta_{j}+D_{j}\beta_{i}-\partial_{t}h_{ij}\right)\,,
\end{eqnarray}
where $D_{i}=e^{\mu}_{i}\nabla_\mu$ is the intrinsic covariant derivative associated with the induced metric defined on the hypersurface, and $e^{\mu}_{i}$ is the appropriate basis vector which is used to transform bulk indices to boundary ones.

Armed with this information, one can write down the Gauss, Codazzi and Ricci
equations, see \cite{Deruelle:2009zk}:
\begin{eqnarray}\label{eq14}
\mathcal{R}_{ijkl}&\equiv&K_{ik}K_{jl}-K_{il}K_{jk}+R_{ijkl}\,, \\ \label{xx1}
\mathcal{R}_{ijk\mathbf{n}}&\equiv&n^{\mu}\mathcal{R}_{ijk\mu}=-D_{i}K_{jk}+D_{j}K_{ik}\,, \\ \label{xxx2}
\mathcal{R}_{i\mathbf{n}j\mathbf{n}}&\equiv&n^{\mu}n^{\nu}\mathcal{R}_{i\mu j\nu}=N^{-1}\big(\partial_{t}K_{ij}-\mathsterling_\beta
K_{ij}\big)+K_{ik}K^{\ k}_{j}+N^{-1}D_i D_jN\,,
\end{eqnarray}
where   in the left hand side of Eq.~(\ref{eq14}) we have the bulk Riemann tensor, but where all indices are now {\it spatial} rather than both spatial and temporal,
and $R_{ijkl}$ is the Riemann tensor constructed purely out of $h_{ij}$, i.e. the metric associated with the hypersurface; and $\pounds_\beta$ is the Lie derivative with respect to shift\footnote{We
have $\pounds_\beta K_{ij}\equiv \beta^{k}D_k K_{ij}+K_{ik}D_j\beta^k+K_{jk}D_i\beta^k$.}.


\subsection{Coframe Slicing}\label{sec4.2}

A key feature of the 3+1 decomposition is the free choice of lapse function and
shift vector which define the choice of foliation at the end. In this paper we stick to the {\it coframe slicing}.
The main advantage for this choice of slicing is the fact that the line element
and therefore components of the infinite derivative function in our gravitational
action will be simplified greatly. In addition, \cite{Anderson:1998cm} has shown that such a slicing has a more transparent form of the canonical action principle and Hamiltonian 
dynamics for gravity. This also leads to a well-posed initial-condition for the evolution of the gravitational constraints in a vacuum by satisfying the Bianchi identities. 
In order to map the ADM line element into the coframe slicing, we use the convention of~\cite{Anderson:1998cm}. We define
\begin{eqnarray} \label{eq15}
\theta^{0}&=&dt\,,\nonumber\\
\theta^{i}&=&dx^{i}+\beta^{i}dt\,,
\end{eqnarray}
where $x^{i}$ and $i=1,2,3$ is the spatial and $t$ is the time coordinates.\footnote{We note that in Eq.~(\ref{eq15}), the $``i"$ for $\theta^{i}$ is just a superscript not a spatial index.}    The metric in the coframe takes the following form
\begin{eqnarray}
\label{eq16}
ds_{\mathrm{coframe}}^{2}=g_{\alpha\beta}\theta^{\alpha}\theta^{\beta}=-N^{2}(\theta^{0})^{2}+g_{ij}\theta^{i}\theta^{j}\,,
\end{eqnarray}
where upon substituting Eq.~(\ref{eq15}) into Eq.~(\ref{eq16})
we recover the original ADM metric given by Eq.~\eqref{eq8}. In this convention, if we
take $\mathbf{g}$ as
the full spacetime metric,
we have the following simplifications:
\begin{eqnarray}\label{eq19}
g_{ij}=h_{ij},\quad g^{ij}=h^{ij},\quad g_{0i}=g^{0i}=0.
\end{eqnarray}
 The convective derivatives $\partial_{\alpha}$ with respect to $\theta^{\alpha}$ are
\begin{align}\label{eq20}
\partial_{0}&\equiv \frac{\partial}{\partial t}-\beta^{i}\partial_{i} \,, \nonumber\\
\partial_{i}&\equiv \frac{\partial}{\partial x^i}\,.
\end{align}
For time-dependent space tensors $T$, we can define the following derivative:
\begin{equation}\label{eq21}
\bar{\partial}_{0}\equiv \frac{\partial}{\partial t}- \pounds_{\beta}\,,
\end{equation}
where $\pounds_{\beta}$ is the Lie derivative with respect to the shift vector $\beta^i$. This is because the off-diagonal components of the coframe metric are zero, {\it i.e.}, $g_{0i}=g^{0i}=0$. 

We shall see later on how this time slicing helps us to simplify the calculations
when the gravitational action contains infinite derivatives.

\subsubsection{Extrinsic Curvature}\label{sec4.2.1}

A change in the choice of time slicing results
in a change of the evolution of the system. The choice of foliation also
has a direct impact on the form of the extrinsic curvature. In this section
we wish to give the form of extrinsic curvature $K_{ij}$ in the coframe slicing. This is due to the fact that the definition of the extrinsic curvature is an initial parameter
that describes the evolution of the system, therefore is it logical for us to derive the extrinsic curvature in the coframe slicing as we use it throughout the paper.
We use \cite{Anderson:1998cm} to find the general definition for $K_{ij}$
in the coframe metric. 
In the coframe, 
\begin{align}\label{eq22}
  {\gamma^\alpha}_{\beta\gamma} &= {\Gamma^\alpha}_{\beta\gamma} + g^{\alpha\delta}
{C^\epsilon}_{\delta(\beta} g_{\gamma)\epsilon} - \frac{1}{2} {C^\alpha}_{\beta\gamma}\,,
\\ \label{eq23}
        d \theta^\alpha &= - \frac{1}{2} {C^\alpha}_{\beta\gamma} \theta^\beta
\wedge \theta^\gamma \,,
\end{align}
where $\Gamma$ is the ordinary Christoffel symbol and ``$\wedge$'' denotes  the exterior or wedge product of vectors $\theta$. By finding the coefficients $C$s 
and subsequently calculating the  connection coefficients $\gamma^{\alpha}_{\beta\gamma}$, one can extract the extrinsic curvature $K_{ij}$ in the coframe setup. We note that the expression
for $d \theta^\alpha$ is the Maurer-Cartan
structure equation~\cite{Hassani}. It is derived from the canonical 1-form $\theta$
on a Lie group $G$ which is the left-invariant $\mathfrak{g}$-valued 1-form
uniquely determined by $\theta(\xi)=\xi$ for all $\xi\in \mathfrak{g}$. 

We
can use differential
forms (See Appendix \ref{A}) to calculate the $C$s,
the coefficients of $d \theta$ where now we can write, 
\begin{eqnarray}\label{eq24}d\theta^{k}=- \left( \partial_i \beta^k \right)
\theta^0
\wedge \theta^i + \frac{1}{2} {C^k}_{ij} \theta^j \wedge \theta^i \,,\end{eqnarray}
where $k=1,2,3$. Now when we insert the $C$s from Appendix \ref{A}, 
\begin{eqnarray}\label{eq25} d \theta^1 = d \left( d x^1 + \beta^1 dt \right)
= d \beta^1
\wedge dt \end{eqnarray}
and 
\begin{eqnarray}\label{eq26}d \theta^0 &=& d(dt) = d^{2}(t) = 0 \nonumber\\
        d \theta^i &=&  d\beta^{i} \wedge dt. \end{eqnarray}
From the definition of $d \theta^\alpha$ in Eq.~(\ref{eq23}) and using the antisymmetric properties
of the $\wedge$ product,
\begin{eqnarray}\label{eq27}
        d \theta^\alpha = - \frac{1}{2} {C^\alpha}_{\beta\gamma} \theta^\beta
\wedge \theta^\gamma 
        = \frac{1}{2} {C^\alpha}_{\beta\gamma} \theta^\gamma
\wedge \theta^\beta = \frac{1}{2} {C^\alpha}_{\gamma\beta} \theta^\beta \wedge \theta^\gamma,
\end{eqnarray}
we get
\begin{eqnarray}\label{eq28}
        {C^\alpha}_{\beta\gamma} =- {C^\alpha}_{\gamma\beta}\,. 
\end{eqnarray}        
Using these properties, we find that ${C^m}_{0i}
= \frac{\partial \beta^m}{\partial x^i}$, ${C^m}_{ij} = 0$ and $C^{0}{}_{ij}=0$. Using Eq.~\eqref{eq22}, we obtain that
\begin{equation}
\gamma_{ij}^{0}=-\frac{1}{2N^2}\Big(h_{il}\partial_{j}(\beta^{l})+h_{jl}\partial_{i}(\beta^{l})-\bar \partial_{0}h_{ij}\Big)\,.
\end{equation} 
Since from Eq.~(\ref{eq13})
\begin{eqnarray}
K_{ij}\equiv - \nabla_{i}n_{j}=\gamma_{ij}^{\mu}n_{\mu}=-N\gamma_{ij}^{0}\,,
\end{eqnarray}
the expression for the extrinsic
curvature in coframe slicing is given by:
\begin{eqnarray}\label{eq29}
K_{ij}=\frac{1}{2N}\Big(h_{il}\partial_{j}(\beta^{l})+h_{jl}\partial_{i}(\beta^{l})-\bar \partial_{0}h_{ij}\Big)\,,
\end{eqnarray}
where $\partial_{0}$ is the time derivative and $\beta^l$ is the ``shift''
in the coframe metric Eq.~(\ref{eq16}).

\subsubsection{Riemann Tensor in the Coframe}

The fact that we move from the ADM metric into the coframe slicing has the following implication on the form of the components of the Riemann tensor. Essentially, since in the coframe slicing in Eq.~(\ref{eq16}) we have $g^{0i}=g_{0i}=0$, therefore we also have, from Eq.~(\ref{eq9}), $n^i=n_i=0$ ($n_{0}$ and $n^0$ stay the same as in Eq.~\eqref{eq9}). Hence the non-vanishing components of the Riemann tensor in the coframe, namely Gauss, Codazzi and Ricci tensor, become: \begin{eqnarray}\label{eq30}
\mathcal{R}_{ijkl}&=&K_{ik}K_{jl}-K_{il}K_{jk}+R_{ijkl}\,,\nonumber \\
\mathcal{R}_{0ijk}&=&N(-D_{k}K_{ji}+D_{j}K_{ki})\,, \nonumber
\\
\mathcal{R}_{0i0j}&=&N(\bar\partial_{0}K_{ij}+NK_{ik}K^{\ k}_{j}+D_i D_jN\,),
\end{eqnarray}
with $\bar\partial_0$ defined in Eq.~(\ref{eq21}) and $D_{j}=e^{\mu}_{j}\nabla_{\mu}$. It can be seen that the Ricci equation, given in Eq.~(\ref{eq14}) is simplified in above due to the definition of Eq.~(\ref{eq21}).
We note that Eq.~(\ref{eq30}) is in the coframe slicing, while Eqs. (\ref{eq14}-\ref{xxx2}) are in the ADM frame only.

\subsubsection{D'Alembertian Operator in Coframe}\label{sec4.2.2}

Since we shall be dealing with a higher-derivative theory of gravity,
it is therefore helpful to first obtain an expression for the $\Box$ operator
in this subsection.
  To do so,
we start off by writing the definition of a single box operator in the coframe, \cite{ChoquetBruhat:1996ak}, 

\begin{eqnarray}\label{boxcoframe}
\Box&=&g^{\mu \nu}\nabla_{\mu}\nabla_{\nu}\nonumber\\
&=& (h^{\mu \nu}+\varepsilon n^{\mu}n^{\nu})\nabla_{\mu}\nabla_{\nu} \nonumber \\
&=&-n^{\mu}n^{\nu}\nabla_{\mu}\nabla_{\nu}+h^{\mu \nu}\nabla_{\mu}\nabla_{\nu}\nonumber\\ &=&-n^{0}n^{0}\nabla_{0}\nabla_{0}+h^{ij}e^{\mu}_{i}e^{\nu
}_{j}\nabla_{\mu}\nabla_{\nu}\nonumber\\
&=& -\frac{1}{N^2}\nabla_{0}\nabla_{0}+h^{ij}D_{i}D_{j} \nonumber \\
&=&-(N^{-1}\bar{\partial_{0}})^{2}+\Box_{hyp}\,,
\end{eqnarray}
where we note that the Greek indices run from $1$ to $4$ and the Latin indices run from
1 to $3$ ($\varepsilon=-1$ for a spacelike hypersurface). We call the spatial box operator $\Box_{hyp}=h^{ij}D_{i}D_{j}$, which stands for
``hypersurface'' as the spatial coordinates are defined on the hypersurface meaning $\Box_{hyp}$ is the projection of the covariant d'Alembertian operator down to the hypersurface, i.e. only the tangential components of the covariant d'Alembertian operator are encapsulated by $\Box_{hyp}$. Also note that in the coframe slicing $g^{ij}=h^{ij}.$ Generalising this result to the $n$th power,  for our purpose, we get 
\begin{eqnarray}\label{4.18}
\mathcal{F}_{i}(\Box)=\sum_{n=0}^{\infty}f_{i_{n}}\left[-(N^{-1}\bar{\partial_{0}})^{2}+\Box_{hyp}\right]^{n}
\end{eqnarray} 
where the $f_{i_{n}}$s are the coefficients of the series.
\section{Generalised  Boundary Term}\label{sec5}

In this section, first we are going to briefly summarise the method of~\cite{Deruelle:2009zk} for finding the boundary term. It has been shown that, given a general gravitational
action 
\begin{eqnarray}\label{5.1}S=\frac{1}{16\pi G}\int_{\mathcal{M}}d^{4}x \, \sqrt{-g}f(\mathcal{R}_{\mu\nu\rho\sigma})\,,
\end{eqnarray}
one can introduce two auxiliary fields $\varrho_{\mu\nu\rho\sigma}$ and $\varphi^{\mu\nu\rho\sigma}$, which are independent of each other and of the metric $g_{\mu\nu}$, while they have all the symmetry properties of the Riemann tensor $\mathcal{R}_{\mu\nu\rho\sigma}$. We can then write down the following equivalent action:
\textcolor{red}{}\begin{eqnarray}\label{5.2}
S=\frac{1}{16\pi G}\int_{\mathcal{M}}d^{4}x \,\sqrt{-g}\left[f(\varrho_{\mu\nu\rho\sigma})
+\varphi^{\mu\nu\rho\sigma}\left(\mathcal{R}_{\mu\nu\rho\sigma}-\varrho_{\mu\nu\rho\sigma}\right)\right]\,.
\end{eqnarray} 
 The reason we introduce these auxiliary fields is that the second derivatives of the metric appear only linearly in~Eq.~\eqref{5.2}. Note that in Eq.~\eqref{5.2}, the terms involving the second derivatives of the metric are not  multiplied by terms of the same type, i.e. involving the second derivative of the metric, so when we integrate by parts once, we are left just with the first derivatives of the metric; we cannot eliminate the first derivatives of the metric as well - since in this paper we are keeping the boundary terms. Note that 
 the first derivatives of the metric are actually contained in these boundary terms if we integrate by parts twice, see our toy model scalar field theory 
 example~in Eqs.~(\ref{eq2},\ref{eq3})~\footnote{This is because $\varrho_{\mu\nu\rho\sigma}$ and $\varphi^{\mu\nu\rho\sigma}$ 
are independent of the metric, and so although $f(\varrho_{\mu\nu\rho\sigma})$ can contain derivatives of $\varrho_{\mu\nu\rho\sigma}$, these are not derivatives of the metric. 
$\mathcal{R}_{\mu\nu\rho\sigma}$ contains a second derivative of the metric but this is the only place where a second derivative of the metric appears in Eq.~\eqref{5.2}}.
Therefore,  terms which are linear in the metric can be eliminated if we integrate by parts; moreover, the use of the auxiliary fields can prove useful in a future Hamiltonian analysis of the action. 

From~\cite{Deruelle:2009zk}, we then decompose the above expression as 
\begin{equation}
\varphi^{\mu\nu\rho\sigma}\left(\mathcal{R}_{\mu\nu\rho\sigma}-\varrho_{\mu\nu\rho\sigma}\right)
=\phi^{ijkl}(\mathcal{R}_{ijkl}-\rho_{ijkl})-4\phi^{ijk}(\mathcal{R}_{ijk\mathbf{n}}-\rho_{ijk})-2\Psi^{ij}(\mathcal{R}_{i\mathbf{n}j\mathbf{n}}-\Omega_{ij})\,,
\end{equation}
where
\begin{eqnarray}\label{tensors1}
\mathcal{R}_{ijkl}\equiv\rho_{ijkl}\equiv\varrho_{ijkl},\qquad \mathcal{R}_{ijk\mathbf{n}}\equiv\rho_{ijk}\equiv n^{\mu}\varrho_{ijk\mu},\qquad\mathcal{R}_{i\mathbf{n}j\mathbf{n}}\equiv\Omega_{ij}\equiv
n^{\mu}n^{\nu}\varrho_{i\mu j\nu}
\end{eqnarray}
are equivalent to the components of the Gauss, Codazzi and Ricci equations given in~Eq.~\eqref{eq14}, also,
\begin{eqnarray}
\phi^{ijkl}\equiv \varphi^{ijkl},\qquad \phi^{ijk}\equiv n_{\mu}\varphi^{ijk\mu},\qquad \Psi^{ij}\equiv -2n_\mu n_\nu\varphi^{i\mu j\nu},
\end{eqnarray}
where $\phi^{ijkl}$, $\phi^{ijk}$ and $\Psi^{ij}$ are spatial tensors evaluated on the hypersurface. 
The equations of motion for the auxiliary fields $\varphi^{\mu \nu \rho \sigma}$ and $\varrho_{\mu \nu \rho \sigma}$ are, respectively given by~\cite{Deruelle:2009zk},
\begin{equation}\label{5.3}
\frac{\delta S}{\delta \varphi^{\mu \nu \rho \sigma}}=0 \Rightarrow \varrho_{\mu\nu\rho\sigma}=\mathcal{R}_{\mu\nu\rho\sigma}\quad \text{and}\quad
\frac{\delta S}{\delta \varrho_{\mu \nu \rho \sigma}}=0 \Rightarrow \varphi^{\mu\nu\rho\sigma}=\frac{\partial f}{\partial\varrho_{\mu\nu\rho\sigma}}\,,
\end{equation}
where $\mathcal{R}_{\mu\nu\rho\sigma}$ is the four-dimensional Riemann tensor. 

One can start from the action given by Eq.~\eqref{5.2}, insert the equation 
of motion for $\varphi^{\mu \nu \rho \sigma}$ and recover the action given by Eq.~\eqref{5.1}. It has been shown by~\cite{Deruelle:2009zk} that one can find the total derivative term of 
the auxiliary action as
\begin{eqnarray}\label{eq5.4}
S=\frac{1}{16 \pi G}\int_{\mathcal{M}}d^{4}x\left( \,
\sqrt{-g}\mathcal{L}-2\partial_{\mu}[\sqrt{-g} \hspace{1mm}
n^{\mu}K \cdot\Psi]\right)\,,
\end{eqnarray}
 where $K=h^{ij}K_{ij}$, with $K_{ij}$ given by Eq.~(\ref{eq29}), and $\Psi=h^{ij}\Psi_{ij}$ , where $\Psi_{ij}$ is given in Eq.~(\ref{5.6}), are spatial tensors evaluated
on the hypersurface $\Sigma_t$ and $\mathcal{L}$ is the Lagrangian density.

In Eq.~(\ref{eq5.4}), the second term is the total derivative. It has been shown that one may add the following action to the above action to eliminate the total derivative appropriately. 
Indeed $\Psi$ can be seen as a modification to the GHY term, which depends on the form of the Lagrangian density~\cite{Deruelle:2009zk}.  
\begin{equation}\label{5.5}
S_{GHY}=\frac{1}{8 \pi G} \oint_{\partial\mathcal{M}}d\Sigma_{\mu}n^{\mu}\Psi \cdot K\,,
\end{equation} 
where
$n^{\mu}$
is the normal vector to the hypersurface and the infinitesimal vector field
\begin{eqnarray}
d\Sigma_{\mu}=\varepsilon_{\mu\alpha\beta\gamma}e^{\alpha}_{1}e^{\beta}_{2}e^{\gamma}_{3}d^{3}y\,,
\end{eqnarray}
is normal to the boundary $\partial\mathcal{M}$ and
is proportional to the volume element of $\partial\mathcal{M}$; in above  $\varepsilon_{\mu\alpha\beta\gamma}=\sqrt{-g}[\mu\,\alpha\,\beta\,\gamma]$
is the Levi-Civita tensor and $y$ are coordinates intrinsic to the boundary~\footnote{We shall also mention that Eq.(\ref{5.5}) is derived from Eq.(\ref{eq5.4}) by performing Stokes theorem, that is $
\int_{\mathcal{M}}A^{\mu}_{\ ;\mu}\sqrt{-g}\hspace{1mm}d^{d}x=\oint_{\partial\mathcal{M}}A^{\mu}\hspace{1mm}d\Sigma_\mu,
$ with $A^{\mu}=n^{\mu}K \cdot\Psi$. }, and we used Eq.~(\ref{eee}). Moreover in Eq.~(\ref{5.5}), we have: 
\begin{equation}\label{5.6}
\Psi^{ij}=-\frac{1}{2}\frac{\delta f}{\delta\Omega_{ij}}\,,
\end{equation}
where $f$ indicates the terms in the Lagrangian density and is built up of tensors $\varrho_{\mu\nu\rho \sigma}$, $\varrho_{\mu\nu}$ and $\varrho  $ as in Eq.~(\ref{5.2}); $G$ is the universal gravitational constant and $\Omega_{ij}$ is given in Eq.~(\ref{tensors1}). Indeed, the above constraint
is extracted from the equation of motion for $\Omega_{ij}$ in the Hamiltonian
regime \cite{Deruelle:2009zk}. In the next section we are going to use the same approach to find the boundary
terms for the most
general, covariant quadratic order action of gravity.
\section{Boundary Terms for Finite Derivative Theory of Gravity}\label{sec6}

In this section we are going to use the 3+1 decomposition and calculate the boundary term of the EH term $\mathcal{R}$, and
$$\mathcal{R}\Box \mathcal{R},~~~\mathcal{R}_{\mu\nu}\Box \mathcal{R}^{\mu\nu},~~~\mathcal{R}_{\mu\nu\rho\sigma}\Box \mathcal{R}^{\mu\nu\rho\sigma},$$ as prescribed in previous section, as a warm-up exercise.

We then move on to our generalised action given in Eq.~(\ref{eq7}).  To decompose any given term, we shall write them in terms of their auxiliary field, therefore we have 
$\mathcal{R}=\varrho,~~~ \mathcal{R}_{\mu\nu}\equiv \varrho_{\mu\nu}$, and $\mathcal{R}_{\mu\nu\rho\sigma}\equiv \varrho_{\mu\nu\rho\sigma}$, 
where the auxiliary fields $\varrho,\varrho_{\mu\nu}$ and  $\varrho_{\mu\nu\rho\sigma}$
have all the symmetry properties of the Riemann tensor. We shall also note that the decomposition of the $\Box$ operator in 3+1 formalism in the coframe 
setup is given by Eq.~(\ref{boxcoframe}). \subsection{$\mathcal{R}$}\label{sec6.1}  
For the Einstein-Hilbert term $\mathcal{R}$, in terms of the auxiliary field $\varrho$ we
find in Appendix \ref{sec:appendixEHterm}
\begin{eqnarray}\label{6.1}
f=\varrho &=& g^{\mu\rho} g^{\nu\sigma} \varrho_{\mu\nu\rho\sigma}\,\nonumber\\
&=& \left( h^{\mu\rho} - n^\mu n^\rho \right) \left( h^{\nu\sigma}
- n^\nu n^\sigma \right) \varrho_{\mu\nu\rho\sigma}\, \nonumber\\
&=& \left( h^{\mu\rho} h^{\nu\sigma} - n^\mu n^\rho h^{\nu\sigma}
- h^{\mu\rho} n^\nu n^\sigma \right) \varrho_{\mu\nu\rho\sigma}\,  \nonumber\\
&=& \left( \rho- 2 \Omega \right),
\end{eqnarray} 
where $\Omega= h^{ij}\Omega_{ij}$ and we used $h^{ij}h^{kl}\rho_{ijkl}=\rho$, and $h^{ij}\rho_{i\nu j\sigma}n^\nu n^\sigma= h^{ij}\Omega_{ij}$ and $\varrho\equiv \mathcal{R}$ in the EH action and the right hand
side of Eq.~(\ref{6.1}) is the $3+1$ decomposed form of the
Lagrangian and hence $\rho$ and $\Omega$ are spatial. We may note that the last term of the expansion on the second line of Eq.~(\ref{6.1}) vanishes due to the symmetry properties of the Riemann tensor.
Using Eq.~(\ref{5.6}), and calculating the functional derivative, we find 
\begin{eqnarray}\label{6.2}\Psi^{ij}=-\frac{1}{2}\frac{\delta f}{\delta\Omega_{ij}}=h^{ij}.\end{eqnarray}
This  verifies the result found in~\cite{Deruelle:2009zk}, and it is clear
that upon substituting this result  into Eq.~(\ref{5.5}), we recover the well known boundary for the
EH action, as $K=h^{ij}K_{ij}$ 
and $\Psi\cdot K\equiv\Psi^{ij}K_{ij}$ where $K_{ij}$ is given by Eq.~(\ref{eq29}).  Hence, 
\begin{equation}\label{GHY-00}
S_{GHY}\equiv S_{0}=\frac{1}{8\pi G}\oint_{\partial\mathcal{M}}d\Sigma_{\mu}n^{\mu} K\,,
\end{equation}
where $d\Sigma_{\mu}$ is the normal to the boundary $\partial\mathcal{M}$ and
is proportional to the volume element of $\partial\mathcal{M}$ while $n^{\mu}$
is the normal vector to the hypersurface.

\subsection{$\mathcal{R}_{\mu\nu\rho\sigma}\Box \mathcal{R}^{\mu\nu\rho\sigma}$}\label{sec6.2}

Next, we start off by writing $\mathcal{R}_{\mu\nu\rho\sigma}\Box \mathcal{R}^{\mu\nu\rho\sigma}$ as its auxiliary equivalent $\varrho_{\mu\nu\rho\sigma}\Box \varrho^{\mu\nu\rho\sigma}$  to obtain 
\begin{eqnarray}\label{eq:decompositionofriemannsquaredintermsofhsandns}
\varrho_{\mu\nu\rho\sigma}\Box \varrho^{\mu\nu\rho\sigma}
&=&\delta_{\mu}^{\alpha}\delta_{\nu}^{\beta}\delta_{\rho}^{\gamma}\delta_{\sigma}^{\lambda}\varrho_{\alpha\beta\gamma\lambda}\Box\varrho^{\mu\nu\rho\sigma}\nonumber\\
&=&\Big[h_{\mu}^{\alpha}h_{\nu}^{\beta}h_{\rho}^{\gamma}h_{\sigma}^{\lambda}-\Big(h_{\mu}^{\alpha}h_{\nu}^{\beta}h_{\rho}^{\gamma}n^{\lambda}n_{\sigma}
+h_{\mu}^{\alpha}h_{\nu}^{\beta}n^{\gamma}n_{\rho}h_{\sigma}^{\lambda}+h_{\mu}^{\alpha}n^{\beta}n_{\nu}h_{\rho}^{\gamma}h_{\sigma}^{\lambda}\nonumber\\
&&+n^{\alpha}n_{\mu}h_{\nu}^{\beta}h_{\rho}^{\gamma}h^\lambda_{\sigma}\Big)+h_{\mu}^{\alpha}n^{\beta}n_{\nu}h_{\rho}^{\gamma}n^{\lambda}n_{\sigma}
+h_{\mu}^{\alpha}n^{\beta}n_{\nu}n^{\gamma}n_{\rho}h_{\sigma}^{\lambda}+n^{\alpha}n_{\mu}h_{\nu}^{\beta}h_{\rho}^{\gamma}n^{\lambda}n_{\sigma}\nonumber\\
&+&n^{\alpha}n_{\mu}h_{\nu}^{\beta}n^{\gamma}n_{\rho}h_{\sigma}^{\lambda}\Big]
\varrho_{\alpha\beta\gamma\lambda}\left(-(N^{-1}\bar{\partial_{0}})^{2}+\Box_{hyp}\right)\varrho^{\mu\nu\rho\sigma}\,,
\end{eqnarray}
where  $\varrho_{\mu\nu\rho\sigma}=\delta_{\mu}^{\alpha}\delta_{\nu}^{\beta}\delta_{\rho}^{\gamma}\delta_{\sigma}^{\lambda}\varrho_{\alpha\beta\gamma\lambda}$ 
(where $\delta^\alpha_\mu$ is the Kronecker delta). This allowed us to use the completeness relation as given in Eq.~(\ref{eq11}). In Eq.~(\ref{eq:decompositionofriemannsquaredintermsofhsandns}), we  used the antisymmetry properties
of the Riemann tensor to eliminate irrelevant terms in the expansion. From Eq.~(\ref{eq:decompositionofriemannsquaredintermsofhsandns}), we
have three types of terms: $$hhhh,~~~~hhhnn,~~~~hhnnnn.$$
The aim is to contract the tensors appearing in Eq.~(\ref{eq:decompositionofriemannsquaredintermsofhsandns}) and extract those terms which are $\Omega_{ij}$ dependent. This is because we only need $\Omega_{ij}$  dependent terms to obtain $\Psi^{ij}$ as in Eq.~(\ref{5.6}) and then the boundary as prescribed in Eq.~(\ref{5.5}). 

A closer look at the expansion given in Eq.~(\ref{eq:decompositionofriemannsquaredintermsofhsandns}) leads us to know which term would admit $\Omega_{ij}$ type terms. Essentially, as defined in Eq.~(\ref{tensors1}), $\Omega_{ij}=n^{\mu}n^{\nu}\varrho_{i\mu j\nu}$, therefore by having two auxiliary field tensors as $\varrho_{\alpha\beta\gamma\lambda}$ and $\varrho^{\mu\nu\rho\sigma}$ in Eq.~(\ref{eq:decompositionofriemannsquaredintermsofhsandns}) (with symmetries of the Riemann tensor) we may construct $\Omega_{ij}$ dependent terms. 
Henceforth, we can see that in this case the $\Omega_{ij}$ dependence comes from the $hhnnnn$ term. 

To see this explicitly, note that in order to perform the appropriate contractions in presence of the d'Alembertian operator, we first need to complete the contractions on the left hand side 
of the $\Box$ operator. We then need to commute the rest of the tensors by using the {\it Leibniz rule} to the right hand side of the components of the operator,  i.e. the $\bar{\partial}_0$'s and the 
$\Box_{hyp}$, and only then do we obtain the $\Omega_{ij}$ type terms. 

We first note that the terms that do not produce $\Omega_{ij}$ dependence are not involved in the boundary calculation, however they might form $\rho_{ijkl}$, $\rho_{ijk}$, 
or their contractions. These terms are equivalent to the Gauss and Codazzi equations as shown in Eq.~(\ref{tensors1}), and we will address their formation in Appendix \ref{RT1}. In addition, as we shall see, by performing the Leibniz rule one produces some associated terms, the $X_{ij}$'s, which appear for example in Eq.~(\ref{ili1}). Again we will keep them only if they are $\Omega_{ij}$ dependent, if not we will drop them.\\

\noindent
$\underline{{\bf hhnnnn}}$ terms: To this end we shall compute the $hhnnnn$ terms, hence we  commute the $h$'s and $n$'s onto the right hand side of the $\Box$
in the $hhnnnn$ term of Eq.~(\ref{eq:decompositionofriemannsquaredintermsofhsandns}):
\begin{eqnarray}\label{ili1}
&&h_{\mu}^{\alpha}n^{\beta}n_{\nu}h_{\rho}^{\gamma}n^{\lambda}n_{\sigma}\varrho_{\alpha\beta\gamma\lambda}\left(-(N^{-1}\bar{\partial_{0}})^{2}+\Box_{hyp}\right)\varrho^{\mu\nu\rho\sigma}\nonumber\\
&&=\left(h_{x}^{i}e_{i}^{\alpha}e_{\mu}^{x}\right)n^{\beta}n_{\nu}\left(h_{y}^{j}e_{j}^{\gamma}e_{\rho}^{y}\right)n^{\lambda}n_{\sigma}\varrho_{\alpha\beta\gamma\lambda}\left(-(N^{-1}\bar{\partial_{0}})^{2}+\Box_{hyp}\right)\varrho^{\mu\nu\rho\sigma}\nonumber\\
&&=\left(h_{x}^{i}e_{\mu}^{x}\right)n_{\nu}\left(h_{y}^{j}e_{\rho}^{y}\right)n_{\sigma}\Omega_{ij}\left(-(N^{-1}\bar{\partial_{0}})^{2}+\Box_{hyp}\right)\varrho^{\mu\nu\rho\sigma}\nonumber\\
&&=-N^{-2}\Omega_{ij}\Big\{\bar{\partial_{0}^{2}}\left(\Omega^{ij}\right)\nonumber\\
&&-\bar{\partial_{0}}\left[\varrho^{\mu\nu\rho\sigma}\bar{\partial_{0}}\left(\left[\left(h_{x}^{i}e_{\mu}^{x}\right)n_{\nu}\left(h_{y}^{j}e_{\rho}^{y}\right)n_{\sigma}\right]\right)\right]-\bar{\partial_{0}}\left(\left[\left(h_{x}^{i}e_{\mu}^{x}\right)
n_{\nu}\left(h_{y}^{j}e_{\rho}^{y}\right)n_{\sigma}\right]\right)\bar{\partial_{0}}\left(\varrho^{\mu\nu\rho\sigma}\right)\Big\}\nonumber\\
&&+\Omega_{ij}\Big\{\Box_{hyp}\left[\Omega^{ij}\right]-D_{a}\left(D^{a}\left[e_{\mu}^{x}n_{\nu}e_{\rho}^{y}n_{\sigma}\right]h_{x}^{i}h_{y}^{j}\varrho^{\mu\nu\rho\sigma}\right)-D_{a}\left[e_{\mu}^{x}n_{\nu}e_{\rho}^{y}n_{\sigma}\right]D^{a}\left(h_{x}^{i}h_{y}^{j}\varrho^{\mu\nu\rho\sigma}\right)\Big\}\nonumber\\
&&=\Omega_{ij}\big(-(N^{-1}\bar{\partial_{0}})^{2}+\Box_{hyp}\big)\Omega^{ij}+\Omega_{ij} X_{1}^{ij}\nonumber\\
&& =\Omega_{ij}\Box\Omega^{ij}+\Omega_{ij} X_{1}^{ij}
\end{eqnarray}

where $\Omega_{ij}\equiv h_{ik} e_{\kappa}^{k}
h_{jm} e_{\lambda}^{m}n_\gamma n_\delta\varrho^{\gamma\kappa\delta\lambda}=h_{ik} h_{jm} n_\gamma n_\delta\varrho^{\gamma k\delta m}$; we note that $X_{1}^{ij}$ only appears 
because of the presence of the $\Box$ operator. \textcolor{red}{}
\begin{eqnarray}\label{x1}
X_{1}^{ij}&=&N^{-2}(\bar{\partial_{0}}\left[\varrho^{\mu\nu\rho\sigma}\bar{\partial_{0}}\left(\left[\left(h_{x}^{i}e_{\mu}^{x}\right)n_{\nu}\left(h_{y}^{j}e_{\rho}^{y}\right)n_{\sigma}\right]\right)\right]+\bar{\partial_{0}}\left(\left[\left(h_{x}^{i}e_{\mu}^{x}\right)
n_{\nu}\left(h_{y}^{j}e_{\rho}^{y}\right)n_{\sigma}\right]\right)\bar{\partial_{0}}\left(\varrho^{\mu\nu\rho\sigma}\right))
\nonumber\\
& -&D_{a}\left(D^{a}\left[e_{\mu}^{x}n_{\nu}e_{\rho}^{y}n_{\sigma}\right]h_{x}^{i}h_{y}^{j}\varrho^{\mu\nu\rho\sigma}\right)-D_{a}\left[e_{\mu}^{x}n_{\nu}e_{\rho}^{y}n_{\sigma}\right]D^{a}\left(h_{x}^{i}h_{y}^{j}\varrho^{\mu\nu\rho\sigma}\right)
 \,.
\end{eqnarray}
The term $\Omega_{rs} X_{1}^{rs}$ will yield $X_{1}^{ij}$ when functionally differentiated with respect to $\Omega_{ij}$ as in Eq.~(\ref{5.6}). Also note $X_{1}^{ij}$ 
does not have any $\Omega^{ij}$ dependence. Similarly for the other $X$ terms which appear later in the paper.
We shall note that when we take $\Box=1$ in Eq.~(\ref{eq:decompositionofriemannsquaredintermsofhsandns}), we obtain,\begin{eqnarray}
&&h_{\mu}^{\alpha}n^{\beta}n_{\nu}h_{\rho}^{\gamma}n^{\lambda}n_{\sigma}\varrho_{\alpha\beta\gamma\lambda}\varrho^{\mu\nu\rho\sigma}\nonumber\\ &=&\left(h_{x}^{i}e_{i}^{\alpha}e_{\mu}^{x}\right)n^{\beta}n_{\nu}\left(h_{y}^{j}e_{j}^{\gamma}e_{\rho}^{y}\right)n^{\lambda}n_{\sigma}\varrho_{\alpha\beta\gamma\lambda}\varrho^{\mu\nu\rho\sigma}\nonumber\\
&=&\left(h_{x}^{i}e_{\mu}^{x}\right)n_{\nu}\left(h_{y}^{j}e_{\rho}^{y}\right)n_{\sigma}\Omega_{ij}\varrho^{\mu\nu\rho\sigma}\nonumber\\
&=&\Omega_{ij}\left(h_{x}^{i}e_{\mu}^{x}\right)n_{\nu}\left(h_{y}^{j}e_{\rho}^{y}\right)n_{\sigma}\varrho^{\mu\nu\rho\sigma}\nonumber\\
&=&\Omega_{ij}\Omega^{ij}\,,
\end{eqnarray}
where we just contract the indices and we do  \textbf{not} need to use the Leibniz rule as we can commute any of the tensors, therefore we do not produce any $X^{ij}$ 
terms at all~\footnote{This is the same for $\Box^2$ and $\Box^n$.}.   
Finally, one can decompose Eq.~(\ref{eq:decompositionofriemannsquaredintermsofhsandns}) as 
\begin{eqnarray}\label{dec1}\varrho_{\mu\nu\rho\sigma}\Box \varrho^{\mu\nu\rho\sigma}=4\Omega_{ij}\Box\Omega^{ij}+4\Omega_{ij} X_{1}^{ij} + \cdots \,,
\end{eqnarray}
where ``$\cdots$'' are terms such as $\rho_{ijkl}\Box\rho^{ijkl}$, $\rho_{ijk}\Box\rho^{ijk}$ and terms that are not $\Omega^{ij}$ dependent and are the results of performing
the Leibniz rule (see Appendix \ref{RT1}). 
When we take $M^2 \to \infty$,
 \textit{i.e.}, when we set $\Box \rightarrow 0$ (recall that $\Box$ has an associated mass scale $\Box/M^2$), which is also  equivalent to considering
$\alpha \rightarrow 0$ in Eq.~(\ref{eq7}), we recover the EH result.

When $\Box \rightarrow 1$, we recover the result for $\mathcal{R}_{\mu\nu\rho\sigma}\mathcal{R}^{\mu\nu\rho\sigma}$ found in~\cite{Deruelle:2009zk}. At both limits, $\Box \rightarrow 0$
and $\Box \rightarrow 1$, the $X_{1}^{ij}$ term is not present. To find the boundary term, we use Eq.~(\ref{5.6}) and then Eq.~(\ref{5.5}). 
We are going to use the Euler-Lagrange equation and drop the total derivatives as a result. We have, 
\begin{eqnarray}
\Psi_{\mathrm{Riem}}^{ij}&=&-\frac{1}{2}\frac{\delta f}{\delta\Omega_{ij}}=-\frac{4}{2}\frac{\delta 
(\Omega_{ij}\Box\Omega^{ij}+\Omega_{ij} X_{1}^{ij})}{\delta \Omega_{ij}}\nonumber\\&=&-2\Bigg\{\frac{\partial
(\Omega_{ij}\Box\Omega^{ij})}{\partial\Omega_{ij}}+\Box\left(\frac{\partial(\Omega_{ij}\Box\Omega^{ij})}{\partial
(\Box\Omega_{ij})}\right)+\frac{\partial (\Omega_{ij} X_{1}^{ij})}{\partial \Omega_{ij}}\Bigg\}\nonumber\\
&=&-2(\Box\Omega^{ij}+\Box \Omega^{ij}+X_{1}^{ij})=-4\Box\Omega^{ij}-2X_{1}^{ij}\,.
\end{eqnarray}
Hence the boundary term for $\mathcal{R}_{\mu\nu\rho\sigma}\Box \mathcal{R}^{\mu\nu\rho\sigma}$ is, 
\begin{equation}\label{eq:finalresultforriemannboxriemann}
S_{1}=-\frac{1}{4\pi G}\oint_{\partial\mathcal{M}}d\Sigma_{\mu}n^{\mu} K_{ij}(2\Box\Omega^{ij}+X_{1}^{ij}).
\end{equation}
where $K_{ij}$\ is given by Eq.~(\ref{eq29}).


\subsection{$\mathcal{R}_{\mu\nu}\Box \mathcal{R}^{\mu\nu}$}\label{sec6.3}

We start by first performing the 3+1 decomposition of $\mathcal{R}_{\mu\nu}\Box\mathcal{R}^{\mu\nu}$ in its auxiliary form $\varrho_{\mu\nu} \Box \varrho^{\mu\nu}$, 
\begin{eqnarray}\label{eq:varrhodecomposition}
&&\varrho_{\mu\nu} \Box \varrho^{\mu\nu}
= g^{\rho\sigma} \varrho_{\rho\mu\sigma\nu} \Box g^{\mu\kappa}
g^{\nu\lambda} g^{\gamma\delta} \varrho_{\gamma\kappa\delta\lambda}\nonumber\\
&&=\left(h^{\rho\sigma} - n^\rho n^\sigma \right) \left( h^{\mu\kappa}
- n^\mu n^\kappa \right) \left( h^{\nu\lambda} - n^\nu n^\lambda \right)
\left( h^{\gamma\delta} - n^\gamma n^\delta \right)  \varrho_{\rho\mu\sigma\nu}
\Box\varrho_{\gamma\kappa\delta\lambda}\nonumber\\
&&=\Big[h^{\rho\sigma} h^{\mu\kappa} h^{\nu\lambda} h^{\gamma\delta}
- \left( n^\rho n^\sigma h^{\mu\kappa} h^{\nu\lambda} h^{\gamma\delta} +
h^{\rho\sigma} n^\mu n^\kappa h^{\nu\lambda} h^{\gamma\delta} 
        + h^{\rho\sigma} h^{\mu\kappa} n^\nu n^\lambda h^{\gamma\delta} +
h^{\rho\sigma} h^{\mu\kappa} h^{\nu\lambda} n^\gamma n^\delta \right) \nonumber\\
        && + n^\rho n^\sigma h^{\mu\kappa} h^{\nu\lambda} n^\gamma n^\delta
+ h^{\rho\sigma} n^\mu n^\kappa n^\nu n^\lambda h^{\gamma\delta} \Big] 
\varrho_{\rho\mu\sigma\nu} \Box \varrho_{\gamma\kappa\delta\lambda}\,,
\end{eqnarray}
where we have used appropriate contractions to write the Ricci tensor in terms of the Riemann tensor. As before, we then used the completeness relation 
Eq.~(\ref{eq11}) and used the 
antisymmetric properties of the Riemann tensor to drop the vanishing terms. We are now set to calculate each term, which we do in more detail in Appendix \ref{RT2}.
Again our aim is to find the $\Omega_{ij}$ dependent terms, by looking at the expansion given in Eq.~(\ref{eq:varrhodecomposition}) and the distribution of the indices, 
the reader can see that the terms which are $\Omega_{ij}$ dependent are those terms which have \textit{at least two} $nn$s contracted with one of the $\varrho$s 
such that we form $n^{\mu}n^{\nu}\varrho_{i\mu j\nu}$.   
\begin{itemize}
\item
{\underline{\bf $hhhnn $}} terms: We start with the $hhhnn $ terms in Eq.~(\ref{eq:varrhodecomposition}). We calculate the first of these in terms of $\Omega_{ik}$ 
and $\rho^{ik}$ also by moving the `$h$'s and `$n$'s onto the right hand side of the $\Box$, 
\begin{eqnarray}
&&n^\rho n^\sigma h^{\mu\kappa} h^{\nu\lambda} h^{\gamma\delta}\varrho_{\rho\mu\sigma\nu}
\left( - \left( N^{-1} \bar{\partial}_0 \right)^2
+ \Box_{hyp} \right) \varrho_{\gamma\kappa\delta\lambda}\nonumber\\
&&=n^\rho n^\sigma (h^{ij} e^{\mu}_{i}e^{\kappa}_{j})(h^{kl} e^{\nu}_{k}e^{\lambda}_{l})(h^{mn}e^{\gamma}_{m}e^{\delta}_{n})\varrho_{\rho\mu\sigma\nu}\left(
- \left( N^{-1} \bar{\partial}_0 \right)^2
+ \Box_{hyp} \right) \varrho_{\gamma\kappa\delta\lambda}\nonumber\\
&&=n^\rho n^\sigma (h^{ij} e^{\kappa}_{j})(h^{kl}
e^{\lambda}_{l})(h^{mn}e^{\gamma}_{m}e^{\delta}_{n})\varrho_{\rho i\sigma
k}\left(
- \left( N^{-1} \bar{\partial}_0 \right)^2
+ \Box_{hyp} \right) \varrho_{\gamma\kappa\delta\lambda}\nonumber\\
&&= \Omega_{ik}(h^{ij} e^{\kappa}_{j})(h^{kl}
e^{\lambda}_{l})(h^{mn}e^{\gamma}_{m}e^{\delta}_{n})\left(
- \left( N^{-1} \bar{\partial}_0 \right)^2
+ \Box_{hyp} \right) \varrho_{\gamma\kappa\delta\lambda}\nonumber\\
&&=-N^{-2}\Omega_{ik}\Big\{\bar{\partial}^{2}_{0}(\rho^{ik})-\bar{\partial}_{0}\big(\varrho_{\gamma\kappa\delta\lambda}\bar{\partial}_{0}[h^{ij}
e^{\kappa}_{j}h^{kl}
e^{\lambda}_{l}h^{mn}e^{\gamma}_{m}e^{\delta}_{n}]\big)-\bar{\partial}_{0}[h^{ij}
e^{\kappa}_{j}h^{kl}
e^{\lambda}_{l}h^{mn}e^{\gamma}_{m}e^{\delta}_{n}]\bar{\partial}_{0}\varrho_{\gamma\kappa\delta\lambda}\Big\}\nonumber\\
&&+\Omega_{ik}\Big\{\Box_{hyp}(\rho^{ik})-D_a\big(\varrho_{\gamma\kappa\delta\lambda}D^a[h^{ij}
e^{\kappa}_{j}h^{kl}
e^{\lambda}_{l}h^{mn}e^{\gamma}_{m}e^{\delta}_{n}]\big)-D_{a}[h^{ij} e^{\kappa}_{j}h^{kl}
e^{\lambda}_{l}h^{mn}e^{\gamma}_{m}e^{\delta}_{n}]D^a\varrho_{\gamma\kappa\delta\lambda}\Big\}\nonumber\\
&&=\Omega_{ik}\Box\rho^{ik}+\Omega_{ik}X_{2(a)}^{ik}\,,
\end{eqnarray}
where the contraction is $h^{ij} e^{\kappa}_{j}h^{kl}
e^{\lambda}_{l}h^{mn}e^{\gamma}_{m}e^{\delta}_{n}\varrho_{\gamma\kappa\delta\lambda}=h^{ij}
h^{kl}
\rho_{jl}=\rho^{ik}$, and
\begin{align}\label{x2a}
X_{2(a)}^{ik}&=N^{-2}\Big\{\bar{\partial}_{0}\big(\varrho_{\gamma\kappa\delta\lambda}\bar{\partial}_{0}[h^{ij}
e^{\kappa}_{j}h^{kl}
e^{\lambda}_{l}h^{mn}e^{\gamma}_{m}e^{\delta}_{n}]\big)+\bar{\partial}_{0}[h^{ij}
e^{\kappa}_{j}h^{kl}
e^{\lambda}_{l}h^{mn}e^{\gamma}_{m}e^{\delta}_{n}]\bar{\partial}_{0}\varrho_{\gamma\kappa\delta\lambda} \Big\}  \nonumber \\
& -D_a\big(\varrho_{\gamma\kappa\delta\lambda}D^a[h^{ij}
e^{\kappa}_{j}h^{kl}
e^{\lambda}_{l}h^{mn}e^{\gamma}_{m}e^{\delta}_{n}]\big)-D_{a}[h^{ij} e^{\kappa}_{j}h^{kl}
e^{\lambda}_{l}h^{mn}e^{\gamma}_{m}e^{\delta}_{n}]D^a\varrho_{\gamma\kappa\delta\lambda} \,.
\end{align}
\item
{\underline{\bf $hhhnn$}} trems: The next $hhhnn$ term in Eq.~(\ref{eq:varrhodecomposition}) is
\begin{eqnarray}
&&h^{\rho\sigma} h^{\mu\kappa} h^{\nu\lambda} n^\gamma n^\delta\varrho_{\rho\mu\sigma\nu}
\left( - \left( N^{-1} \bar{\partial}_0 \right)^2
+ \Box_{hyp} \right) \varrho_{\gamma\kappa\delta\lambda}\nonumber\\
&&=(h^{ij} e^{\rho}_{i}e^{\sigma}_{j})(h^{kl} e^{\mu}_{k}e^{\kappa}_{l})(
h^{mn} e^{\nu}_{m}e^{\lambda}_{n})n^\gamma n^\delta\varrho_{\rho\mu\sigma\nu}
\left( - \left( N^{-1} \bar{\partial}_0 \right)^2
+ \Box_{hyp} \right) \varrho_{\gamma\kappa\delta\lambda}\nonumber\\
&&=\rho_{ k m}(h^{kl} e^{\kappa}_{l})(
h^{mn} e^{\lambda}_{n})n^\gamma n^\delta
\left( - \left( N^{-1} \bar{\partial}_0 \right)^2
+ \Box_{hyp} \right) \varrho_{\gamma\kappa\delta\lambda}\nonumber\\
&&=-N^{-2}\rho_{ k m}\Big\{\bar{\partial}^{2}_{0}(\Omega^{km})-\bar{\partial}_{0}\big(\varrho_{\gamma\kappa\delta\lambda}\bar{\partial}_{0}[h^{kl}
e^{\kappa}_{l}
h^{mn} e^{\lambda}_{n}n^\gamma n^\delta]\big)-\bar{\partial}_{0}[h^{kl} e^{\kappa}_{l}
h^{mn} e^{\lambda}_{n}n^\gamma n^\delta]\bar{\partial}_{0}\varrho_{\gamma\kappa\delta\lambda}\Big\}\nonumber\\
&&+\rho_{ k m}\Big\{\Box_{hyp}(\Omega^{km})-D_a\big(\varrho_{\gamma\kappa\delta\lambda}D^a[h^{kl}
e^{\kappa}_{l}
h^{mn} e^{\lambda}_{n}n^\gamma n^\delta]\big)-D_{a}[h^{kl} e^{\kappa}_{l}
h^{mn} e^{\lambda}_{n}n^\gamma n^\delta]D^a\varrho_{\gamma\kappa\delta\lambda}\Big\} \nonumber\\
&& =\rho_{ k m}\Box\Omega^{ k m}+\cdots\,,
\end{eqnarray}
where we used $h^{kl} e^{\kappa}_{l}
h^{mn} e^{\lambda}_{n}n^\gamma n^\delta\varrho_{\gamma\kappa\delta\lambda}=h^{kl}
h^{mn} n^\gamma n^\delta\varrho_{\gamma l\delta n}=\Omega^{km}$ and we note that ``$\cdots$'' are extra terms which do not depend on $\Omega^{km}$. 

\item
{\underline{\bf $hhnnnn$}} terms: The the next term in Eq.~(\ref{eq:varrhodecomposition}) is of the form $hhnnnn$:
\begin{eqnarray}
&& n^\rho n^\sigma h^{\mu\kappa} h^{\nu\lambda} n^\gamma n^\delta\varrho_{\rho\mu\sigma\nu}
\left( - \left( N^{-1} \bar{\partial}_0 \right)^2
+ \Box_{hyp} \right) \varrho_{\gamma\kappa\delta\lambda}\nonumber\\
&& =n^\rho n^\sigma (h^{ij} e^{\mu}_{i}e^{\kappa}_{j})(
h^{kl} e^{\nu}_{k}e^{\lambda}_{l} )n^\gamma n^\delta\varrho_{\rho\mu\sigma\nu}
\left( - \left( N^{-1} \bar{\partial}_0 \right)^2
+ \Box_{hyp} \right) \varrho_{\gamma\kappa\delta\lambda}\nonumber\\
&& =  \Omega^{jl}e^{\kappa}_{j}e^{\lambda}_{l} n^\gamma n^\delta
\left( - \left( N^{-1} \bar{\partial}_0 \right)^2
+ \Box_{hyp} \right) \varrho_{\gamma\kappa\delta\lambda}\nonumber\\
&&=-N^{-2}\Omega^{jl}\Big\{\bar{\partial}^{2}_{0}(\Omega_{jl})-\bar{\partial}_{0}\big(\varrho_{\gamma\kappa\delta\lambda}\bar{\partial}_{0}[e^{\kappa}_{j}e^{\lambda}_{l}
n^\gamma n^\delta]\big)-\bar{\partial}_{0}[e^{\kappa}_{j}e^{\lambda}_{l}
n^\gamma n^\delta]\bar{\partial}_{0}\varrho_{\gamma\kappa\delta\lambda}\Big\}\nonumber\\
&&+\Omega^{jl}\Big\{\Box_{hyp}(\Omega_{jl})-D_a\big(\varrho_{\gamma\kappa\delta\lambda}D^a[e^{\kappa}_{j}e^{\lambda}_{l}
n^\gamma n^\delta]\big)-D_{a}[e^{\kappa}_{j}e^{\lambda}_{l}
n^\gamma n^\delta]D^a\varrho_{\gamma\kappa\delta\lambda}\Big\} \nonumber\\
&& =\Omega^{jl}\Box\Omega_{jl}+\Omega^{jl}X_{2(b)jl}\,,
\end{eqnarray}
where $e^{\kappa}_{j}e^{\lambda}_{l}
n^\gamma n^\delta\varrho_{\gamma\kappa\delta\lambda}=n^\gamma n^\delta\varrho_{\gamma
j\delta l}=\Omega_{jl}$, and
\begin{align}\label{x2b}
X_{2(b)jl}&=N^{-2}\Big\{\bar{\partial}_{0}\big(\varrho_{\gamma\kappa\delta\lambda}\bar{\partial}_{0}[e^{\kappa}_{j}e^{\lambda}_{l}
n^\gamma n^\delta]\big)+\bar{\partial}_{0}[e^{\kappa}_{j}e^{\lambda}_{l}
n^\gamma n^\delta]\bar{\partial}_{0}\varrho_{\gamma\kappa\delta\lambda} \Big\} \nonumber \\
& -D_a\big(\varrho_{\gamma\kappa\delta\lambda}D^a[e^{\kappa}_{j}e^{\lambda}_{l}
n^\gamma n^\delta]\big)-D_{a}[e^{\kappa}_{j}e^{\lambda}_{l}
n^\gamma n^\delta]D^a\varrho_{\gamma\kappa\delta\lambda} \,.
\end{align}
\item
{\underline{\bf $hhnnnn$}} terms: Finally, the  last  $hhnnnn$ terms in Eq.~(\ref{eq:varrhodecomposition}) is
\begin{eqnarray}
&& h^{\rho\sigma} n^\mu n^\kappa n^\nu n^\lambda h^{\gamma\delta}\varrho_{\rho\mu\sigma\nu}
\left( - \left( N^{-1} \bar{\partial}_0 \right)^2
+ \Box_{hyp} \right) \varrho_{\gamma\kappa\delta\lambda}\nonumber\\
&&=(h^{ij}e^{\rho}_{i}e^{\sigma}_{j}) n^\mu n^\kappa n^\nu n^\lambda (h^{mn}e^{\gamma}_{m}e^{\delta}_{n})\varrho_{\rho\mu\sigma\nu}
\left( - \left( N^{-1} \bar{\partial}_0 \right)^2
+ \Box_{hyp} \right) \varrho_{\gamma\kappa\delta\lambda}\nonumber\\
&&= \Omega n^\kappa n^\lambda h^{mn}e^{\gamma}_{m}e^{\delta}_{n}
\left( - \left( N^{-1} \bar{\partial}_0 \right)^2
+ \Box_{hyp} \right) \varrho_{\gamma\kappa\delta\lambda}\nonumber\\
&&=-N^{-2}\Omega\Big\{\bar{\partial}^{2}_{0}(\Omega)-\bar{\partial}_{0}\big(\varrho_{\gamma\kappa\delta\lambda}\bar{\partial}_{0}[n^\kappa
n^\lambda h^{mn}e^{\gamma}_{m}e^{\delta}_{n}]\big)-\bar{\partial}_{0}[n^\kappa
n^\lambda h^{mn}e^{\gamma}_{m}e^{\delta}_{n}]\bar{\partial}_{0}\varrho_{\gamma\kappa\delta\lambda}\Big\}\nonumber\\
&&+\Omega\Big\{\Box_{hyp}(\Omega)-D_a\big(\varrho_{\gamma\kappa\delta\lambda}D^a[n^\kappa
n^\lambda h^{mn}e^{\gamma}_{m}e^{\delta}_{n}]\big)-D_{a}[n^\kappa n^\lambda
h^{mn}e^{\gamma}_{m}e^{\delta}_{n}]D^a\varrho_{\gamma\kappa\delta\lambda}\Big\} \nonumber\\
&& =\Omega\Box\Omega+\Omega X_{2(c)}\,,
\end{eqnarray}
\end{itemize}
where we used 
$ n^\kappa n^\lambda h^{mn}e^{\gamma}_{m}e^{\delta}_{n}\varrho_{\gamma\kappa\delta\lambda}=
h^{mn}\Omega_{m n}=\Omega$, and
\begin{align}\label{x2c}
X_{2(c)}&= N^{-2} \Big\{\bar{\partial}_{0}\big(\varrho_{\gamma\kappa\delta\lambda}\bar{\partial}_{0}[n^\kappa
n^\lambda h^{mn}e^{\gamma}_{m}e^{\delta}_{n}]\big)+\bar{\partial}_{0}[n^\kappa
n^\lambda h^{mn}e^{\gamma}_{m}e^{\delta}_{n}]\bar{\partial}_{0}\varrho_{\gamma\kappa\delta\lambda} \Big\} \nonumber \\
& -D_a\big(\varrho_{\gamma\kappa\delta\lambda}D^a[n^\kappa
n^\lambda h^{mn}e^{\gamma}_{m}e^{\delta}_{n}]\big)-D_{a}[n^\kappa n^\lambda
h^{mn}e^{\gamma}_{m}e^{\delta}_{n}]D^a\varrho_{\gamma\kappa\delta\lambda} \,.
\end{align}
Summarising this result, we can write Eq.~(\ref{eq:varrhodecomposition}), as
\begin{eqnarray}\varrho_{\mu\nu} \Box \varrho^{\mu\nu}&=& \Omega(\Box
\Omega+X_{2(c)})+ \Omega_{ij}( \Box \Omega^{ij}+X_{2(b)}^{ij}) - \rho_{ij} \Box
\Omega^{ij}\nonumber\\&-& \Omega_{ij}( \Box \rho^{ij}+X_{2(a)}^{ij})+\cdots \,,
\end{eqnarray}
where ``$\cdots$'' are the contractions of $\rho_{ijkl}$ and $\rho_{ijk}$ (see Appendix \ref{RT2}) and  the terms that are the results of performing Leibniz rule,
which have no $\Omega_{ij}$ dependence. When $\Box \rightarrow 1$, we recover the result for $\mathcal{R}_{\mu\nu}\mathcal{R}^{\mu\nu}$ found in~\cite{Deruelle:2009zk}. 
At both limits, $\Box \rightarrow 0$ and $\Box \rightarrow 1$, the $X_{2}$ terms are not present. 
Obtaining the boundary term requires us to extract $\Psi^{ij}$ as it is given  in Eq.~(\ref{5.6}). 
Hence the boundary for $\mathcal{R}_{\mu\nu}\Box \mathcal{R}^{\mu\nu}$ is given by,  
 \begin{align}\label{eq:finalresultforriccitensorboxriccitensor}
S_{2}&=- \frac{1}{8\pi
G}\oint_{\partial\mathcal{M}}d\Sigma_{\mu}n^{\mu}\Big[
K\Box\Omega+ K_{ij}\Box\Omega^{ij}-K_{ij}\Box\rho^{ij}\Big]\nonumber \\
&-\frac{1}{16\pi G}\oint_{\partial\mathcal{M}}d\Sigma_{\mu}n^{\mu}\Big[K X_{2(c)}+K_{ij}(X_{2(b)}^{ij}-X_{2(a)}^{ij}) \Big]\,,
\end{align}
where $K\equiv h^{ij} K_{ij}$ and $K_{ij}$ is given by Eq.~(\ref{eq29}).


\subsection{$\mathcal{R}\Box \mathcal{R}$}\label{sec6.4}

We do not need to commute any $h$'s, or $n$'s across the $\Box$ here, we can simply apply Eq.~(\ref{6.1}) to $\varrho \Box \varrho$, the auxiliary equivalent of the $\mathcal{R}\Box \mathcal{R}$ 
term:
\begin{eqnarray}\label{36}
\varrho\Box \varrho = \left( \rho - 2 \Omega \right) \Box
\left( \rho - 2 \Omega \right)\,,
\end{eqnarray}
whereupon extracting $\Psi^{ij}$ using Eq.~(\ref{5.6}), and using Eq.~(\ref{5.5}) as in the previous cases, we obtain the boundary term for $\mathcal{R}\Box \mathcal{R}$ to be
\begin{eqnarray}\label{eq:finalresultforrboxr}
S_{3}=- \frac{1}{4\pi
G}\oint_{\partial\mathcal{M}}d\Sigma_{\mu} \, n^{\mu}\Big[
2K\Box \Omega -K\Box \rho \Big],
\end{eqnarray}
where $K\equiv h^{ij} K_{ij}$ and $K_{ij}$ is given by  Eq.~(\ref{eq29}). Again   when $\Box \rightarrow 1$, we recover the result for $\mathcal{R}^2$ found in~\cite{Deruelle:2009zk}.

\subsection{Full result} 

Summarising the results of Eq.~(\ref{eq:finalresultforriemannboxriemann}), Eq.~(\ref{eq:finalresultforriccitensorboxriccitensor}) and Eq.~(\ref{eq:finalresultforrboxr}), altogether we have
\begin{align}\label{eq:finalresultforallboundarytermswithonebox}
S
&=\frac{1}{16\pi G}\int _{\mathcal{M}}d^{4}x \, \sqrt{-g}\Big[\varrho+\alpha\big(\varrho\Box \varrho+\varrho_{\mu\nu}\Box \varrho^{\mu\nu}+\varrho_{\mu\nu\rho\sigma}\Box \varrho^{\mu\nu\rho\sigma}\big)+\varphi^{\mu\nu\rho\sigma}\left(\mathcal{R}_{\mu\nu\rho\sigma}-\varrho_{\mu\nu\rho\sigma}\right)\Big]\nonumber \\
&-\frac{1}{8\pi G}\oint_{\partial\mathcal{M}}d\Sigma_{\mu} \, n^{\mu}\Big[-K+\alpha\big(-2K\Box \varrho+4K\Box\Omega+K\Box \Omega+4K_{ij}\Box\Omega^{ij}-K_{ij}\Box\rho^{ij}+K_{ij}\Box\Omega^{ij}\big)\Big]\nonumber \\
&-\frac{1}{16\pi G}\oint_{\partial\mathcal{M}}d\Sigma_{\mu}n^{\mu}\alpha\Big[K X_{2(c)}+K_{ij}(4X_{1}^{ij}+X_{2(b)}^{ij}-X_{2(a)}^{ij}) \Big] \nonumber \\
&=\frac{1}{16\pi G}\int _{\mathcal{M}}d^{4}x \, \sqrt{-g}\Big[\varrho+\alpha\big(\varrho\Box
\varrho+\varrho_{\mu\nu}\Box \varrho^{\mu\nu}+\varrho_{\mu\nu\rho\sigma}\Box \varrho^{\mu\nu\rho\sigma}\big)+\varphi^{\mu\nu\rho\sigma}\left(\mathcal{R}_{\mu\nu\rho\sigma}-\varrho_{\mu\nu\rho\sigma}\right)\Big]\nonumber \\
&-\frac{1}{8\pi G}\oint_{\partial\mathcal{M}}d\Sigma_{\mu} \, n^{\mu}\Big[-K+\alpha\big(-2K\Box\rho+5K\Box\Omega+5K_{ij}\Box\Omega^{ij}-K_{ij}\Box\rho^{ij}\Big] \nonumber \\
& -\frac{1}{16\pi G}\oint_{\partial\mathcal{M}}d\Sigma_{\mu}n^{\mu}\alpha\Big[K X_{2(c)}+K_{ij}(4X_{1}^{ij}+X_{2(b)}^{ij}-X_{2(a)}^{ij}) \Big]\,. 
\end{align}
 This result matches with the EH action~\cite{Deruelle:2009zk}, when we take the limit $\Box \rightarrow 0$; that is, we are left with the same expression for boundary 
 as in Eq.~(\ref{GHY-00}):
\begin{align}
S_{EH}&=\frac{1}{16\pi G}\int _{\mathcal{M}}d^{4}x \, \sqrt{-g}\Big[\varrho+\varphi^{\mu\nu\rho\sigma}\left(\mathcal{R}_{\mu\nu\rho\sigma}-\varrho_{\mu\nu\rho\sigma}\right)\Big]\nonumber \\
&+\frac{1}{8\pi G}\oint_{\partial\mathcal{M}}d\Sigma_{\mu} \, n^{\mu}K \,, \end{align}
since the $X$-type terms are not present when $\Box \rightarrow 0$.
When $\Box \rightarrow 1$, we recover the result for $\mathcal{R}+\alpha(\mathcal{R}^{2}+\mathcal{R}_{\mu \nu}\mathcal{R}^{\mu \nu}+\mathcal{R}_{\mu\nu\rho\sigma}\mathcal{R}^{\mu\nu\rho\sigma})$ found in~\cite{Deruelle:2009zk}; that is, we are left with
\begin{align}
S &= \frac{1}{16\pi G}\int _{\mathcal{M}}d^{4}x \, \sqrt{-g}\Big[\varrho+\alpha\big(\varrho^{2}+\varrho_{\mu\nu} \varrho^{\mu\nu}+\varrho_{\mu\nu\rho\sigma} \varrho^{\mu\nu\rho\sigma}\big)+\varphi^{\mu\nu\rho\sigma}\left(\mathcal{R}_{\mu\nu\rho\sigma}-\varrho_{\mu\nu\rho\sigma}\right)\Big]\nonumber \\
&-\frac{1}{8\pi G}\oint_{\partial\mathcal{M}}d\Sigma_{\mu} \, n^{\mu}\Big[-K+\alpha\big(-2K\rho+5K\Omega+5K_{ij}\Omega^{ij}-K_{ij}\rho^{ij}\Big]\,; \end{align}
again the $X$-type terms are not present when $\Box \rightarrow 1$.
We should note that the $X_1$ and $X_2$ terms are the results of having the covariant  d'Alembertian operator so, in the absence of the d'Alembertian operator, one does not produce them at all and hence the result found
in~\cite{Deruelle:2009zk} is guaranteed.
\\\\
We may now turn our attention to the $\mathcal{R}\Box^{2}
\mathcal{R},\mathcal{R}_{\mu\nu}\Box^{2} \mathcal{R}^{\mu\nu}$ and $\mathcal{R}_{\mu\nu\rho\sigma}\Box^{2} \mathcal{R}^{\mu\nu\rho\sigma}$. 
Here the methodology will remain the same. One first decomposes each term into its $3+1$ equivalent.   
Then one extracts $\Psi^{ij}$ using Eq.~(\ref{5.6}), and then the boundary terms can be obtained using Eq.~(\ref{5.5}). In this case we will have two operators, namely
\begin{eqnarray}\Box^2=\left( - \left( N^{-1}
\bar{\partial}_0 \right)^2
+ \Box_{hyp} \right)\left( - \left( N^{-1} \bar{\partial}_0 \right)^2
+ \Box_{hyp} \right)
\end{eqnarray} 
This means that upon expanding to $3+1$, one performs the Leibniz rule twice and hence obtains eight total derivatives that do not produce any 
$\Omega_{ij}$s or its contractions that are relevant to the boundary calculations and hence must be dropped.


\subsection{Generalisation to Infinite Derivative Theory of Gravity}\label{sec6.5}

We may now turn our attention to the infinite
derivative terms; namely, $\mathcal{R}\mathcal{F}_{1}(\Box)\mathcal{R},~~\mathcal{R}_{\mu\nu}\mathcal{F}_{2}(\Box)\mathcal{R}^{\mu\nu}$ 
and $\mathcal{R}_{\mu\nu\rho\sigma}\mathcal{F}_{3}(\Box)\mathcal{R}^{\mu\nu\rho\sigma}$. For such cases, we can write down the following relation
(see Appendix~\ref{B4}):
\begin{eqnarray}\label{33}
&&X D^{2n} Y= D^{2n}(XY)-D^{2n-1}(D(X)Y)-D^{2n-2}(D(X)D(Y))\nonumber \\
&&-D^{2n-3}(D(X)D^{2}(Y))-\dots-D(D(X)D^{2n-2}(Y))-D(X)D^{2n-1}(Y)\,,
\end{eqnarray}
where $X$ and $Y$ are tensorial structures such as $\varrho_{\mu\nu\rho\sigma}$, $\varrho_{\mu\nu}$, $\varrho$ and their contractions, while $D$ denotes any operators. 
These operators do not have to be differential operators and indeed this result can be generalised to cover the case where there are different types of operator and a similar (albeit more complicated) structure is recovered.

From (\ref{33}), one produces $2n$ total derivatives, analogous to the scalar toy model case, see Eqs.~(\ref{eq2},\ref{eq3}). We can then write the $3+1$ decompositions for each curvature by generalising  Eq.~(\ref{eq:finalresultforriemannboxriemann}), ~Eq.~(\ref{eq:finalresultforriccitensorboxriccitensor})
and Eq.~(\ref{eq:finalresultforrboxr}) and writing $\mathcal{R}_{\mu\nu\rho\sigma} F_3(\Box) \mathcal{R}^{\mu\nu\rho\sigma}$, $\mathcal{R}_{\mu\nu} F_2(\Box) \mathcal{R}^{\mu\nu}$ and $\mathcal{R} F_1(\Box) \mathcal{R}$ in terms of their auxiliary equivalents  $\varrho_{\mu\nu\rho\sigma} F_3(\Box) \varrho^{\mu\nu\rho\sigma}$,
$\varrho_{\mu\nu} F_2(\Box) \varrho^{\mu\nu}$ and $\varrho F_1(\Box)
\varrho$.
Then 
\begin{eqnarray}
\varrho_{\mu\nu\rho\sigma}\mathcal{F}_{3}(\Box) \varrho^{\mu\nu\rho\sigma}=4\Omega_{ij}\mathcal{F}_{3}(\Box)\Omega^{ij}+4\Omega_{ij}X_{1}^{ij}+\cdots \,,
\end{eqnarray}
\begin{eqnarray}\label{x2}
        \varrho_{\mu\nu} \mathcal{F}_{2}(\Box) \varrho^{\mu\nu}&=& \Omega\mathcal{F}_{2}(\Box)
\Omega+ \Omega_{ij} \mathcal{F}_{2}(\Box) \Omega^{ij} - \rho_{ij} \mathcal{F}_{2}(\Box)
\Omega^{ij}\nonumber\\&-& \Omega_{ij} \mathcal{F}_{2}(\Box) \rho^{ij}+\Omega_{ij}X_{2}^{ij}+\cdots \,,
\end{eqnarray}
\begin{eqnarray}\label{x3}\varrho\mathcal{F}_{1}(\Box) \varrho = \left( \rho
- 2 \Omega \right) \mathcal{F}_{1}(\Box)
\left( \rho - 2 \Omega \right)\,,
\end{eqnarray}
where we have dropped the irrelevant terms, as we did before, while $X_{1}^{ij}$ and $X_{2}^{ij}$ are the analogues of Eqs.~\eqref{x1},~\eqref{x2a},~\eqref{x2b},~\eqref{x2c}. We now need to use the generalised form of the Euler-Lagrange equations to obtain the $\Psi^{ij}$ in each case:
\begin{align}\label{37}
\frac{\delta f}{\delta \Omega_{ij}}&=\frac{\partial
f}{\partial
\Omega_{ij}}-\nabla_{\mu}\left(\frac{\partial
f}{\partial
(\nabla_{\mu}\Omega_{ij})} \right)+\nabla_{\mu}\nabla_{\nu}\left(\frac{\partial
f}{\partial
(\nabla_{\mu}\nabla_{\nu}\Omega_{ij})} \right)+\cdots \nonumber\\
&=\frac{\partial
f}{\partial
\Omega_{ij}}+\sum^{\infty}_{n=1}\Box^{n}\left(\frac{\partial
f}{\partial (\Box^{n}\Omega_{ij})}\right)\,,
\end{align}
where we have imposed that $\delta \Omega_{ij}=0$ on the boundary $\partial \mathcal{M}$.

Hence, by using $\Omega =h^{ij} \Omega_{ij}$ and $\rho=h^{ij} \rho_{ij}$,  we find in Appendix~\ref{C} that:
\begin{eqnarray}\label{functionaldifferentation}
\frac{\delta \big(\Omega\mathcal{F}(\Box)\Omega\big)}{\delta \Omega_{ij}}=2h^{ij}\mathcal{F}(\Box)\Omega,&\quad&
\frac{\delta \big(\Omega_{ij}\mathcal{F}(\Box)\Omega^{ij}\big)}{\delta \Omega_{ij}}=2\mathcal{F}(\Box)\Omega^{ij}\nonumber\\
\frac{\delta \big(\rho\mathcal{F}(\Box)\Omega\big)}{\delta
\Omega_{ij}}=h^{ij}\mathcal{F}(\Box)\rho,&\quad& \frac{\delta \big(\rho_{ij}\mathcal{F}(\Box)\Omega^{ij}\big)}{\delta
\Omega_{ij}}=\mathcal{F}(\Box)\rho^{ij}\nonumber\\
\frac{\delta \big(\Omega\mathcal{F}(\Box)\rho\big)}{\delta
\Omega_{ij}}=h^{ij}\mathcal{F}(\Box)\rho,&\quad&
\frac{\delta \big(\Omega_{ij}\mathcal{F}(\Box)\rho^{ij}\big)}{\delta
\Omega_{ij}}=\mathcal{F}(\Box)\rho^{ij}\,,
\end{eqnarray}
and so using Eq.~(\ref{5.6}), the $\Psi^{ij}$s are:
\begin{eqnarray}\label{39}
\Psi^{ij}_{\mathrm{Riem}}&=&-4\mathcal{F}_{3}(\Box)\Omega^{ij}-2X_{1}^{ij}\nonumber\\
\Psi^{ij}_{\mathrm{Ric}}&=&\mathcal{F}_{2}(\Box)\rho^{ij}-h^{ij}\mathcal{F}_{2}(\Box)\Omega-\mathcal{F}_{2}(\Box)\Omega^{ij}-\frac{1}{2}X_{2}^{ij}\nonumber\\
\Psi^{ij}_{\mathrm{Scal}}&=&2h^{ij}\mathcal{F}_{1}(\Box)\big(-2\Omega+\rho\big)\equiv 2 h^{ij}\mathcal{F}_{1}(\Box)\varrho\,,
\end{eqnarray}
where we have used Eq.~(\ref{6.1}) in the last line. Finally, we can use Eq.~\eqref{5.5} and write the boundary terms corresponding to our infinite-derivative action as, 
\begin{align}\label{40}
S_{tot}&=S_{gravity}+S_{boundary}\nonumber \\&=\frac{1}{16\pi G}\int_{\mathcal{M}}
d^{4}x \, \sqrt{-g}\Big[\varrho+\alpha\big(\varrho\mathcal{F}_{1}(\Box)\varrho+\varrho_{\mu\nu}\mathcal{F}_{2}(\Box)\varrho^{\mu\nu}\nonumber \\
&+\varrho_{\mu\nu\rho\sigma}\mathcal{F}_{3}(\Box)\varrho^{\mu\nu\rho\sigma}\big)+\varphi^{\mu\nu\rho\sigma}\left(\mathcal{R}_{\mu\nu\rho\sigma}-\varrho_{\mu\nu\rho\sigma}\right)\Big]\nonumber \\
& +\frac{1}{8\pi
G}\oint_{\partial\mathcal{M}}d\Sigma_{\mu} \, n^{\mu}\Big[K+\alpha\big(2K\mathcal{F}_{1}(\Box)\rho-4K\mathcal{F}_{1}(\Box)\Omega \nonumber \\
&-K\mathcal{F}_{2}(\Box)\Omega- K_{ij}\mathcal{F}_{2}(\Box)\Omega^{ij}+K_{ij}
\mathcal{F}_{2}(\Box)\rho^{ij}-4K_{ij}\mathcal{F}_{3}(\Box)\Omega^{ij}-2X_{1}^{ij}-\frac{1}{2}X_{2}^{ij}\big)\Big]\,.
\end{align}
where $\Omega_{ij}= n^\gamma n^\delta\varrho_{\gamma i\delta j}$, $\Omega=h^{ij} \Omega_{ij}$, $\rho_{ij}=h^{km} \rho_{ijkm}$, $\rho=h^{ij} \rho_{ij}$, $K=h^{ij} K_{ij}$ 
and $K_{ij}$ is the extrinsic curvature given by Eq.~(\ref{eq29}). 
We note that when we decompose the $\Box$, after we perform the Leibniz rule enough times, we can reconstruct the $\Box$ in its original form, i.e. it is not affected by the use of the coframe. In this way, we can always reconstruct ${\cal F}_{i}(\Box)$. However, the form of the $X$-type terms will depend on the decomposition and therefore the use of the coframe. 
In this regard, the $X$-type terms depend on the coframe but the ${\cal F}_{i} (\Box)$ terms do not.

\section{Conclusion}

This paper generalises earlier contributions for finding the boundary term for a higher derivative theory of gravity. Our work has focused 
on seeking the boundary term or GHY contribution for a covariant infinite derivative theory of gravity, which is quadratic in curvature.
Such modifications are inevitable for any covariant construction of gravitational theory. In stringy parlance, such infinite derivative 
quadratic order curvature correction entails to one-loop in string coupling $g_s$, and all order $\alpha'$ corrections. Indeed, here one 
can possibly make the connection even stronger, if one could find an exact form of the form factors, i.e. ${\cal F}(\Box)$'s from closed-string field theory. 
However, note 
that our considerations are much more general than any such possible stringy correction, as the analytical construction does not rely either
on supersymmetry or in any specific spacetime 
dimensions. Indeed, we only concentrated on pure {\it massless } gravitational degrees of freedom, which is also in stark contrast 
with the string theory set-up.

Indeed, in this case some novel features distinctively filter through our analysis. Since the bulk action contains nonlocal form factors,
${\cal F}_{i} (\Box)$, the boundary action also contains the nonlocality, as can be seen from our final expression Eq.~(\ref{40}).
The above expression {\it also} has a smooth limit when $M\rightarrow \infty$, or $\Box \rightarrow 0$, which is the local limit of Eq.~(\ref{choice-a}), and our 
results then reproduce the GHY term, and when ${\cal F}_{i}(\Box) \rightarrow 1$, our results coincide with that of~\cite{Deruelle:2009zk}. 

Our study will have implications for finding the Hamiltonian for an infinite derivative theory of gravity, as well as seeking the entanglement entropy, and 
to understand gravity/field theory correspondence in Anti (de)Sitter spacetimes. Some of these issues
will be studied in future publication.

\acknowledgments
The authors would like to thank Tirthabir Biswas, Alex Koshelev, Terry Tomboulis,  and Valery Frolov for discussions. AM would like to thank
Kapteyn Institute for Astrophysics, Casa Mathem\'atica Oaxaca (CMO), and UC Berkley Gump 
station, French Polynesia, Moorea, for their kind hospitality, where part of this research was conducted. The work of A.M. is supported in part by the Lancaster-Manchester-Sheffield Consortium for Fundamental 
Physics under STFC grant ST/L000520/1. ST is supported by a scholarship from the Onassis Foundation.

\appendix
\section{$K_{ij}$ in the Coframe Metric}\label{A}

In this section we wish to use the approach of~\cite{Anderson:1998cm} and
find the general definition for $K_{ij}$ in the coframe metric. 
Given,
\begin{eqnarray} \label{eq:defitionofextrinsiccurvaturegivenbyArlen}
  {\gamma^\alpha}_{\beta\gamma} &=& {\Gamma^\alpha}_{\beta\gamma} + g^{\alpha\delta}
{C^\epsilon}_{\delta(\beta} g_{\gamma)\epsilon} - \frac{1}{2} {C^\alpha}_{\beta\gamma}\,,
\\ \label{zuzu}
        d \theta^\alpha &=& - \frac{1}{2} {C^\alpha}_{\beta\gamma} \theta^\beta
\wedge \theta^\gamma \,,
\end{eqnarray}
where $\Gamma$ is the ordinary Christoffel symbol, $\wedge$ is the ordinary wedge product and the $C$s are coefficients
to be found. 
By comparing the values given in \cite{Anderson:1998cm} with the ordinary Christoffel symbols, we
can see that 
\begin{eqnarray}\label{eq:differentCs}
{C^i}_{00} = {C^0}_{0i} = {C^0}_{i0} = {C^0}_{00} = 0\,, \nonumber \\
        {C^i}_{0k} = {C^i}_{k0} + 2 \partial_k \beta^i\,, \nonumber \\
        {C^i}_{jk} = {{C_k}^i}_{j} + {{C_j}^i}_{k}\,, 
\end{eqnarray}
Now in the coframe metric in Eq.~(\ref{eq16}),  
\begin{eqnarray}
 && g^{0\delta} {C^\epsilon}_{\delta(i} g_{j)\epsilon} \,, \nonumber \\
        &=& - \frac{1}{N^2} \left[ {C^\epsilon}_{0(i} g_{j)\epsilon} \right]\,,
\nonumber \\
        &=& - \frac{1}{2N^2} \left[ {C^\epsilon}_{0i} g_{j\epsilon} + {C^\epsilon}_{0j}
g_{i\epsilon} \right]\,, \nonumber \\
        &=& - \frac{1}{2N^2} \left[ {C^m}_{0i} g_{jm} + {C^m}_{0j} g_{im}
\right] \,,\nonumber \\
&=& -\frac{1}{2N^2} \left[C_{j0i}+C_{i0j} \right]
\end{eqnarray}
and $C_{j0i} = g_{\alpha j} {C^\alpha}_{0i}=g_{jk}{C^k}_{0i}$. 

In general, for a $p$-form $\alpha$ and a $q$-form $\beta,$
\begin{align}
\alpha \wedge \beta &= (-1)^{pq} \beta \wedge \alpha \,, \\
d(\alpha \wedge \beta) &= (d \alpha)\wedge \beta + (-1)^{p}\alpha \wedge (d \beta) \,.
\end{align}
Hence, if $p$ is odd,
\begin{eqnarray} \label{eq:antisymmetricpropertiesofwedgeproduct}
\alpha \wedge \alpha = (-1)^{p^2} \alpha \wedge \alpha =-\alpha
\wedge \alpha= 0 \,.
\end{eqnarray}
From Eq.~(\ref{zuzu}) and Eq.~(\ref{eq:differentCs}) we can see that 
\begin{eqnarray}
d \theta^1 &=& - \frac{1}{2} {C^1}_{\beta\gamma} \theta^\beta \wedge \theta^\gamma\,,
 \nonumber\\
        &=& - \frac{1}{2} {C^1}_{0i} \theta^0 \wedge \theta^i - \frac{1}{2}
{C^1}_{i0} \theta^i \wedge \theta^0 - \frac{1}{2} {C^1}_{ij} \theta^i \wedge
\theta^j \,,\nonumber\\
        &=& - \frac{1}{2} \left[ {C^1}_{i0} + 2 \partial_i \beta^1 \right]
\theta^0 \wedge \theta^i + \frac{1}{2} {C^1}_{i0} \theta^0 \wedge \theta^i+
\frac{1}{2} {C^1}_{ij} \theta^j \wedge \theta^i \,, \nonumber\\
        &=& - \left( \partial_i \beta^1 \right) \theta^0 \wedge \theta^i
+ \frac{1}{2} {C^1}_{ij} \theta^j \wedge \theta^i\,.
\end{eqnarray}
We get a similar result for $d\theta^2$ and $d\theta^3$, so we can say that
\begin{eqnarray}\label{eq:definitionofdthetak}
d\theta^{k}=- \left( \partial_i \beta^k \right) \theta^0
\wedge \theta^i + \frac{1}{2} {C^k}_{ij} \theta^j \wedge \theta^i \,,
\end{eqnarray}
where $k=1,2,3$. Now from the definition of $\theta$ in Eq.~(\ref{eq23}), 
\begin{eqnarray}\label{eq:defntheta1} 
d \theta^1 = d \left( d x^1 + \beta^1 dt \right) = d \beta^1
\wedge dt \,,
\end{eqnarray}
and 
\begin{eqnarray}\label{eq:dthetazerovanishesanddefnofdthetai}
        d \theta^0 &=& d(dt) = d^{2}(t) = 0 \,,\nonumber\\
        d \theta^i &=& d \left( dx^i + \beta^i dt \right)\,, \nonumber\\
        &=& d \left(dx^i \right) + d \left( \beta^i \wedge dt \right) \,,\nonumber\\
        &=& d \left( \beta^i \wedge dt \right)\,, \nonumber \\
        &=& d\beta^{i} \wedge dt \,.
        \end{eqnarray}
        Let us point out that $\beta^{i}dt=\beta^{i} \wedge dt$. 
\begin{eqnarray} \theta^0 \wedge \theta^i &=& dt \wedge \left( dx^i + \beta^i
\wedge dt \right) \,, \nonumber\\
&=& dt \wedge dx^i \nonumber\\
        \theta^i \wedge \theta^j &=& \left( dx^i + \beta^i dt \right) \wedge
\left( d x^j + \beta^j dt \right) \,, \nonumber\\
        &=& dx^i \wedge dx^j + dx^i \wedge \left( \beta^j dt \right) + \left(
\beta^i dt \right) \wedge dx^j + \left( \beta^i dt \right) \wedge \left(
\beta^j \wedge dt \right) \,,\nonumber\\
        &=& dx^i \wedge dx^j + dx^i \wedge \beta^j \wedge dt + \beta^i \wedge
dt  \wedge dx^j \,,\nonumber\\
        &=& dx^i \wedge dx^j + \beta^j \wedge dx^i \wedge dt - \beta^i \wedge
dx^j \wedge dt\,.
\end{eqnarray}
Now using Eq.~(\ref{eq:definitionofdthetak}) and  Eq.~(\ref{eq:dthetazerovanishesanddefnofdthetai}), 
\begin{eqnarray} \label{eq:dthetakintermsofwedge}
d \theta^k &=& d \beta^k \wedge dt  \,,\nonumber\\
        &=& \left( \frac{\partial \beta^k}{\partial x^1} dx^1 + \frac{\partial
\beta^k}{\partial x^2} dx^2 + \frac{\partial \beta^k}{\partial x^3} dx^3
\right) \wedge dt \,,\nonumber\\
        &=& - \left( \partial_i \beta^k \right) dt \wedge dx^i - \frac{1}{2}
{C^k}_{ij} \left[ dx^i \wedge dx^j + \beta^j \wedge dx^i \wedge dt - \beta^i
\wedge dx^j \wedge dt \right] \,,\end{eqnarray}
where $k=1,2,3$. 
From the definition of $d \theta^\alpha$ in Eq.~(\ref{zuzu}) and using the antisymmetric properties
of the $\wedge$ product from Eq.~(\ref{eq:antisymmetricpropertiesofwedgeproduct}),
\begin{eqnarray}
        d \theta^\alpha &=& - \frac{1}{2} {C^\alpha}_{\beta\gamma} \theta^\beta
\wedge \theta^\gamma \,,\nonumber\\
        &=& - \frac{1}{2} {C^\alpha}_{\gamma\beta} \theta^\gamma \wedge \theta^\beta\,,
\nonumber\\
        &=& \frac{1}{2} {C^\alpha}_{\gamma\beta} \theta^\beta \wedge \theta^\gamma\,,
\end{eqnarray}
and therefore
\begin{eqnarray}\label{eq:symmetryofC}
        {C^\alpha}_{\beta\gamma} = -{C^\alpha}_{\gamma\beta} \,,
\end{eqnarray}
we can then write
\begin{eqnarray}
        && {C^\alpha}_{00} = {C^\alpha}_{11} = {C^\alpha}_{22} = {C^\alpha}_{33}
= 0 \,,\nonumber\\
        && {C^\alpha}_{0i} = - {C^\alpha}_{i0} \,.
\end{eqnarray}
Combining Eq.~(\ref{zuzu}), Eq.~(\ref{eq:dthetazerovanishesanddefnofdthetai}), Eq.~(\ref{eq:dthetakintermsofwedge})
and utilising Eq.~(\ref{eq:symmetryofC})
\begin{eqnarray}
        0 &=& d\theta^0 = - \frac{1}{2} {C^0}_{\beta\gamma} \theta^\gamma
\wedge \theta^\beta \,,\nonumber\\
        &=& - \frac{1}{2} {C^0}_{0i} \theta^0 \wedge \theta^i - \frac{1}{2}
{C^0}_{i0} \theta^i \wedge \theta^0 - \frac{1}{2} {C^0}_{ij} \theta^i \wedge
\theta^j \text{(for } i \neq j) \,,\nonumber\\
        &=& - {C^0}_{0i} \theta^0 \wedge \theta^i - {C^0}_{ij} \theta^i \wedge
\theta^j \text{(for } i < j) \,,\nonumber\\
        &=& - {C^0}_{01} \theta^0 \wedge \theta^1- - {C^0}_{02} \theta^0
\wedge \theta^2 - {C^0}_{03} \theta^0 \wedge \theta^3 
        - {C^0}_{12} \theta^1 \wedge \theta^2 - {C^0}_{13} \theta^1 \wedge
\theta^3 - {C^0}_{23} \theta^2 \wedge \theta^3 \,,\nonumber\\
        &=& - {C^0}_{01} dt \wedge dx^1 - {C^0}_{02} dt \wedge dx^2 - {C^0}_{03}
dt \wedge dx^3 \,, \nonumber\\
        &&- {C^0}_{12} \left[ dx^1 \wedge dx^2 + \beta^2 dx^1 \wedge dt -
\beta^1 dx^2 \wedge dt \right]\,, \nonumber\\
        && - {C^0}_{13} \left[ dx^1 \wedge dx^3 + \beta^3 dx^1 \wedge dt
- \beta^1 dx^3 \wedge dt \right] \,,\nonumber\\
        && - {C^0}_{23} \left[ dx^2 \wedge dx^3 + \beta^3 dx^2 \wedge dt
- \beta^2 dx^3 \wedge dt \right]\,.
\end{eqnarray}   
In order for this to be satisfied,  each term must vanish separately as the $dx^\alpha
\wedge dx^j$ are linearly independent and so the coefficient
of each must be zero
and thus ${C^0}_{12} =  {C^0}_{13} = {C^0}_{23}={C^0}_{01}={C^0}_{02} ={C^0}_{03}=0$
and thus ${C^0}_{\alpha\beta}=0$. Similarly using Eqs.~(\ref{zuzu}), (\ref{eq:dthetazerovanishesanddefnofdthetai}), (\ref{eq:dthetakintermsofwedge}) and (\ref{eq:symmetryofC})
\begin{eqnarray}
        d \beta^1 \wedge dt &=& \frac{\partial \beta^1}{\partial dx^1} dx^1
+  \frac{\partial \beta^1}{\partial dx^2} dx^2 +  \frac{\partial \beta^1}{\partial
dx^3} dx^3 \,,\nonumber\\
        &=& d \theta^1 = - {C^1}_{0i} \theta^0 \wedge \theta^i - {C^1}_{ij}
\theta^i \wedge \theta^j \,,\nonumber\\
        &=& - {C^1}_{01} dt \wedge dx^1 - {C^1}_{02} dt \wedge dx^2 - {C^1}_{03}
dt \wedge dx^3 \,,\nonumber\\
        &&- {C^1}_{12} \left[ dx^1 \wedge dx^2 + \beta^2 dx^1 \wedge dt -
\beta^1 dx^2 \wedge dt \right] \,,\nonumber\\
        && - {C^1}_{13} \left[ dx^1 \wedge dx^3 + \beta^3 dx^1 \wedge dt
- \beta^1 dx^3 \wedge dt \right] \,,\nonumber\\
        && - {C^1}_{23} \left[ dx^2 \wedge dx^3 + \beta^3 dx^2 \wedge dt
- \beta^2 dx^3 \wedge dt \right]\,.
\end{eqnarray}
Again, in order for this relation to be satisfied, ${C^1}_{12} = {C^1}_{13}
= {C^1}_{23} = 0$ and ${C^1}_{01} = \frac{\partial \beta^1}{\partial x^1}$,
${C^1}_{02} = \frac{\partial \beta^1}{\partial x^2}$, ${C^1}_{03} = \frac{\partial
\beta^1}{\partial x^3}$. We deduce that ${C^m}_{0i}
= \frac{\partial \beta^m}{\partial x^i}$, ${C^m}_{ij} = 0$ and $C^{0}{}_{ij}=0$. Using Eq.~\eqref{zuzu} and that in the coframe 
$\Gamma^0_{ij} = \frac{1}{2} \frac{1}{N^2} \bar \partial_0 h_{ij}$, we obtain that
\begin{eqnarray}
\gamma_{ij}^{0}=-\frac{1}{2N^2}\Big(h_{il}\partial_{j}(\beta^{l})+h_{jl}\partial_{i}(\beta^{l})-\bar \partial_{0}h_{ij}\Big)\,.
\end{eqnarray}
Since from Eq.~(\ref{eq13})
\begin{eqnarray}
K_{ij}\equiv - \nabla_{i}n_{j}=\gamma_{ij}^{\mu}n_{\mu}=-N\gamma_{ij}^{0}\,,
\end{eqnarray}
Eq. \eqref{eq:defitionofextrinsiccurvaturegivenbyArlen}
becomes 
\begin{eqnarray}K_{ij}=\frac{1}{2N}(h_{il}\partial_{j}(\beta^{l})+h_{jl}\partial_{i}(\beta^{l})-\bar \partial_{0}h_{ij})\,.\end{eqnarray}

\section{3+1 Decompositions}\label{B}

\subsection{Einstein-Hilbert term}\label{sec:appendixEHterm}

We can write the Einstein-Hilbert term $\mathcal{R}$ as its auxiliary equivalent $\varrho$. Then we can use the completeness relation Eq.~(\ref{eq11}) to show that
\begin{eqnarray}\label{eq:decompositionofthericciscalarauxequiv}
\varrho &=& g^{\mu\rho} g^{\nu\sigma} \varrho_{\mu\nu\rho\sigma} \,,\nonumber\\
        &=& \left( h^{\mu\rho} - n^\mu n^\rho \right) \left( h^{\nu\sigma}
- n^\nu n^\sigma \right) \varrho_{\mu\nu\rho\sigma} \,,\nonumber\\
        &=& \left( h^{\mu\rho} h^{\nu\sigma} - n^\mu n^\rho h^{\nu\sigma}
- h^{\mu\rho} n^\nu n^\sigma + n^\mu n^\rho n^\nu n^\sigma \right) \varrho_{\mu\nu\rho\sigma}\,,
\nonumber\\
        &=& \left( h^{\mu\rho} h^{\nu\sigma} - n^\mu n^\rho h^{\nu\sigma}
- h^{\mu\rho} n^\nu n^\sigma \right) \varrho_{\mu\nu\rho\sigma} \,,\nonumber\\
        &=& \left( \rho - 2 \Omega \right)\,,
\end{eqnarray}
noting that the term with four $n^\alpha$s vanishes due to the antisymmetry
of the Riemann tensor in the first and last pair of indices (recall that $\varrho_{\mu\nu\rho\sigma}$ has the same symmetry properties as the Riemann tensor)
\subsection{Riemann Tensor}\label{RT1}
In this section we wish to show the contraction of the rest of the terms in Eq.~(\ref{eq:decompositionofriemannsquaredintermsofhsandns}) for the sake of completeness. We have, from $hhhh,$ 
\begin{eqnarray}
&&h_{\mu}^{\alpha}h_{\nu}^{\beta}h_{\rho}^{\gamma}h_{\sigma}^{\lambda}\varrho_{\alpha\beta\gamma\lambda}\left[-(N^{-1}\bar{\partial_{0}})^{2}+\Box_{hyp}\right]\varrho^{\mu\nu\rho\sigma}\nonumber\\
&&=\left(h_{\mu}^{i}e_{i}^{\alpha}\right)\left(h_{\nu}^{j}e_{j}^{\beta}\right)\left(h_{\rho}^{k}e_{k}^{\gamma}\right)\left(h_{\sigma}^{l}e_{l}^{\lambda}\right)\varrho_{\alpha\beta\gamma\lambda}\Big[-(N^{-1}\bar{\partial_{0}})^{2}+\Box_{hyp}\Big]\varrho^{\mu\nu\rho\sigma}\nonumber\\
&&=\left(h_{\mu}^{i}\right)\left(h_{\nu}^{j}\right)\left(h_{\rho}^{k}\right)\left(h_{\sigma}^{l}\right)\rho_{ijkl}\Big[-(N^{-1}\bar{\partial_{0}})^{2}+\Box_{hyp}\Big]\varrho^{\mu\nu\rho\sigma}\nonumber\\
&&=-N^{-2}\left[\left(h_{m}^{i}e_{\mu}^{m}\right)\left(h_{n}^{j}e_{\nu}^{n}\right)\left(h_{x}^{k}e_{\rho}^{x}\right)\left(h_{y}^{l}e_{\sigma}^{y}\right)\right]\rho_{ijkl}\Big[-(N^{-1}\bar{\partial_{0}})^{2}+\Box_{hyp}\Big]\varrho^{\mu\nu\rho\sigma}\nonumber\\
&&=-N^{-2}\rho_{ijkl}\Big\{\bar{\partial_{0}^{2}}\left(\rho^{ijkl}\right)\nonumber\\
&&-\bar{\partial_{0}}\left[\varrho^{\mu\nu\rho\sigma}\bar{\partial_{0}}\left(\left[\left(h_{m}^{i}e_{\mu}^{m}\right)\left(h_{n}^{j}e_{\nu}^{n}\right)\left(h_{x}^{k}e_{\rho}^{x}\right)\left(h_{y}^{l}e_{\sigma}^{y}\right)\right]\right)\right]\nonumber\\
&&-\bar{\partial_{0}}\left(\left[\left(h_{m}^{i}e_{\mu}^{m}\right)\left(h_{n}^{j}e_{\nu}^{n}\right)\left(h_{x}^{k}e_{\rho}^{x}\right)\left(h_{y}^{l}e_{\sigma}^{y}\right)\right]\right)\bar{\partial_{0}}\left(\varrho^{\mu\nu\rho\sigma}\right)\Big\}\nonumber\\
&&+\rho_{ijkl}\Big\{\Box_{hyp}\left[\rho^{ijkl}\right]-D_{a}\left(D^{a}\left[e_{\mu}^{m}e_{\nu}^{n}e_{\rho}^{x}e_{\sigma}^{y}\right]h_{m}^{i}h_{n}^{j}h_{x}^{k}h_{y}^{l}\varrho^{\mu\nu\rho\sigma}\right)\nonumber\\
&&-D_{a}\left[e_{\mu}^{m}e_{\nu}^{n}e_{\rho}^{x}e_{\sigma}^{y}\right]D^{a}\left(h_{m}^{i}h_{n}^{j}h_{x}^{k}h_{y}^{l}\varrho^{\mu\nu\rho\sigma}\right)\Big\}
\end{eqnarray}
which produced $\rho_{ijkl}\Box\rho^{ijkl}$ and the terms which are the results of Leibniz rule. Next in Eq.~(\ref{eq:decompositionofriemannsquaredintermsofhsandns}) is, 
\begin{eqnarray}
&&h_{\mu}^{\alpha}h_{\nu}^{\beta}h_{\rho}^{\gamma}n^{\lambda}n_{\sigma}\varrho_{\alpha\beta\gamma\lambda}\left(-(N^{-1}\bar{\partial_{0}})^{2}+\Box_{hyp}\right)\varrho^{\mu\nu\rho\sigma}\nonumber\\
&&=\left(h_{\mu}^{i}e_{i}^{\alpha}\right)\left(h_{\nu}^{j}e_{j}^{\beta}\right)\left(h_{\rho}^{k}e_{k}^{\gamma}\right)n^{\lambda}n_{\sigma}\varrho_{\alpha\beta\gamma\lambda}\left(-(N^{-1}\bar{\partial_{0}})^{2}+\Box_{hyp}\right)\varrho^{\mu\nu\rho\sigma}\nonumber\\
&&=\left(h_{m}^{i}e_{\mu}^{m}\right)\left(h_{n}^{j}e_{\nu}^{n}\right)\left(h_{x}^{k}e_{\rho}^{x}\right)n^{\lambda}n_{\sigma}\varrho_{ijk\lambda}\left(-(N^{-1}\bar{\partial_{0}})^{2}+\Box_{hyp}\right)\varrho^{\mu\nu\rho\sigma}\nonumber\\
&&=\left(h_{m}^{i}e_{\mu}^{m}\right)\left(h_{n}^{j}e_{\nu}^{n}\right)\left(h_{x}^{k}e_{\rho}^{x}\right)n_{\sigma}\rho_{ijk}\left(-(N^{-1}\bar{\partial_{0}})^{2}+\Box_{hyp}\right)\varrho^{\mu\nu\rho\sigma}\nonumber\\
&&=-N^{-2}\rho_{ijk}\Big\{\bar{\partial_{0}^{2}}\left(\rho^{ijk}\right)\nonumber\\
&&-\bar{\partial_{0}}\left[\varrho^{\mu\nu\rho\sigma}\bar{\partial_{0}}\left(\left[h_{m}^{i}e_{\mu}^{m}h_{n}^{j}e_{\nu}^{n}h_{x}^{k}e_{\rho}^{x}n_{\sigma}\right]\right)\right]-\bar{\partial_{0}}\left(\left[h_{m}^{i}e_{\mu}^{m}
h_{n}^{j}e_{\nu}^{n}h_{x}^{k}e_{\rho}^{x}n_{\sigma}\right]\right)\bar{\partial_{0}}\left(\varrho^{\mu\nu\rho\sigma}\right)\Big\}\nonumber\\
&&+\rho_{ijk}\Big\{\Box_{hyp}\left[\rho^{ijk}\right]-D_{a}\left(D^{a}\left[e_{\mu}^{m}e_{\nu}^{n}e_{\rho}^{x}n_{\sigma}\right]h_{m}^{i}h_{n}^{j}h_{x}^{k}\varrho^{\mu\nu\rho\sigma}\right)\nonumber\\
&&-D_{a}\left[e_{\mu}^{m}e_{\nu}^{n}e_{\rho}^{x}n_{\sigma}\right]D^{a}\left(h_{m}^{i}h_{n}^{j}h_{x}^{k}\varrho^{\mu\nu\rho\sigma}\right)\Big\}\end{eqnarray}
with $\rho_{ijk}\equiv n^\mu \rho_{ijk\mu}$. Here we produced $\rho_{ijk}\Box\rho^{ijk}$ and the extra terms which are the results
of the Leibniz rule.
Similarly we can find the contractions for different terms in Eq.~(\ref{eq:decompositionofriemannsquaredintermsofhsandns}). \subsection{Ricci Tensor}\label{RT2}
In similar way as we did in the Riemann case we can find all the other contractions in the expansion of Eq.~(\ref{eq:varrhodecomposition}) which we omitted. They are:
\begin{eqnarray}
&&h^{\rho\sigma} h^{\mu\kappa} h^{\nu\lambda} h^{\gamma\delta}
\varrho_{\rho\mu\sigma\nu} \Box \varrho_{\gamma\kappa\delta\lambda} \nonumber\\
&&=  (h^{im}e^{\rho}_{i}e^{\sigma}_{m})(h^{jn}e^{\mu}_{j}e^{\kappa}_{n})(h^{kx}e^{\nu}_{k}e^{\lambda}_{x})(h^{ly}e^{\gamma}_{l}e^{\delta}_{y})\varrho_{\rho\mu\sigma\nu}
 \left( - \left( N^{-1} \bar{\partial}_0 \right)^2
+ \Box_{hyp} \right) \varrho_{\gamma\kappa\delta\lambda}\nonumber\\
&&= (h^{jn}e^{\kappa}_{n})(h^{kx}e^{\lambda}_{x})(h^{ly}e^{\gamma}_{l}e^{\delta}_{y})\rho_{jk}
 \left( - \left( N^{-1} \bar{\partial}_0 \right)^2
+ \Box_{hyp} \right) \varrho_{\gamma\kappa\delta\lambda}\nonumber\\
&&=-N^{-2}\rho_{jk}\Big\{\bar{\partial}^{2}_{0}(\rho^{jk})-\bar{\partial}_{0}(\varrho_{\gamma\kappa\delta\lambda}\bar{\partial}_{0}[(h^{jn}e^{\kappa}_{n})(h^{kx}e^{\lambda}_{x})(h^{ly}e^{\gamma}_{l}e^{\delta}_{y})])\nonumber\\
&&-\bar{\partial}_{0}[(h^{jn}e^{\kappa}_{n})(h^{kx}e^{\lambda}_{x})(h^{ly}e^{\gamma}_{l}e^{\delta}_{y})]\bar{\partial}_{0}\varrho_{\gamma\kappa\delta\lambda}\Big\}\nonumber\\
&&+\rho_{jk}\Big\{\Box_{hyp}(\rho^{jk})-D_{a}(\varrho_{\gamma\kappa\delta\lambda}D^{a}[(h^{jn}e^{\kappa}_{n})(h^{kx}e^{\lambda}_{x})(h^{ly}e^{\gamma}_{l}e^{\delta}_{y})])\nonumber\\
&&-D_{a}[(h^{jn}e^{\kappa}_{n})(h^{kx}e^{\lambda}_{x})(h^{ly}e^{\gamma}_{l}e^{\delta}_{y})]D^{a}\varrho_{\gamma\kappa\delta\lambda}\Big\}
\end{eqnarray} 
with $(h^{jn}e^{\kappa}_{n})(h^{kx}e^{\lambda}_{x})(h^{ly}e^{\gamma}_{l}e^{\delta}_{y})\varrho_{\gamma\kappa\delta\lambda}=\rho^{jk}$. Above we produced $\rho_{jk}\Box\rho^{jk}$ plus other terms that are results of the Leibniz rule. And, 
\begin{eqnarray}
&&h^{\rho\sigma} n^\mu n^\kappa h^{\nu\lambda} h^{\gamma\delta}\varrho_{\rho\mu\sigma\nu}
\left( - \left( N^{-1} \bar{\partial}_0 \right)^2
+ \Box_{hyp} \right) \varrho_{\gamma\kappa\delta\lambda}\nonumber\\
&&= (h^{ij} e^{\rho}_{i}e^{\sigma}_{j})n^\mu n^\kappa (h^{kl} e^{\nu}_{k}e^{\lambda}_{l})(h^{mn}e^{\gamma}_{m}e^{\delta}_{n})\varrho_{\rho\mu\sigma\nu}
\left( - \left( N^{-1} \bar{\partial}_0 \right)^2
+ \Box_{hyp} \right) \varrho_{\gamma\kappa\delta\lambda}\nonumber\\
&&=  n^\kappa (h^{kl} e^{\lambda}_{l})(h^{mn}e^{\gamma}_{m}e^{\delta}_{n})\rho_{
k}
\left( - \left( N^{-1} \bar{\partial}_0 \right)^2
+ \Box_{hyp} \right) \varrho_{\gamma\kappa\delta\lambda}\nonumber\\
&&=-N^{-2}\rho_{
k}\Big\{\bar{\partial}^{2}_{0}(\rho^{k})-\bar{\partial}_{0}\big(\varrho_{\gamma\kappa\delta\lambda}\bar{\partial}_{0}[n^\kappa
h^{kl} e^{\lambda}_{l}h^{mn}e^{\gamma}_{m}e^{\delta}_{n}]\big)-\bar{\partial}_{0}[n^\kappa
h^{kl} e^{\lambda}_{l}h^{mn}e^{\gamma}_{m}e^{\delta}_{n}]\bar{\partial}_{0}\varrho_{\gamma\kappa\delta\lambda}\Big\}\nonumber\\
&&+\rho_{
k}\Big\{\Box_{hyp}(\rho^{k})-D_a\big(\varrho_{\gamma\kappa\delta\lambda}D^a[n^\kappa
h^{kl} e^{\lambda}_{l}h^{mn}e^{\gamma}_{m}e^{\delta}_{n}]\big)-D_{a}[n^\kappa
h^{kl} e^{\lambda}_{l}h^{mn}e^{\gamma}_{m}e^{\delta}_{n}]D^a\varrho_{\gamma\kappa\delta\lambda}\Big\}\nonumber\\
\end{eqnarray}
where we used $n^\mu\varrho_{\mu k}=\rho_k$ and $n^\kappa h^{kl} e^{\lambda}_{l}h^{mn}e^{\gamma}_{m}e^{\delta}_{n}\varrho_{\gamma\kappa\delta\lambda}=n^\kappa
h^{kl} \varrho_{\kappa l}=\rho^k$.
We produced $\rho_{k}\Box\rho^{k}$ plus other terms that are results
of the Leibniz rule.
We may also note that one can write, $\rho_{ij}\equiv h^{kl}\rho_{ikjl}$, $\rho\equiv h^{ik}h^{il}\rho_{ijkl}$
and $\rho_{i}\equiv h^{jk}\rho_{jik}$.
\subsection{Generalisation from $\Box$ to $\mathcal{F}(\Box)$}\label{B4}
In Eq.~(\ref{ili1}) for $\Box^2$, we have, 
\begin{eqnarray}
&&\Omega_{ij}\left[h_{x}^{i}e_{\mu}^{x}n_{\nu}h_{y}^{j}e_{\rho}^{y}n_{\sigma}\right]\Box^{2}\varrho^{\mu\nu\rho\sigma}\nonumber\\
&&=\Omega_{ij}\left[h_{x}^{i}e_{\mu}^{x}n_{\nu}h_{y}^{j}e_{\rho}^{y}n_{\sigma}\right]\left(-(N^{-1}\bar{\partial_{0}})^{2}+\Box_{hyp}\right)\left(-(N^{-1}\bar{\partial_{0}})^{2}+\Box_{hyp}\right)\varrho^{\mu\nu\rho\sigma}\,,\nonumber\\
&&=N^{-4}\Omega_{ij}\left[h_{x}^{i}e_{\mu}^{x}n_{\nu}h_{y}^{j}e_{\rho}^{y}n_{\sigma}\right]\bar{\partial_{0}^{4}}\varrho^{\mu\nu\rho\sigma}\nonumber\\
&&-N^{-2}\Omega_{ij}\left[h_{x}^{i}e_{\mu}^{x}n_{\nu}h_{y}^{j}e_{\rho}^{y}n_{\sigma}\right]\bar{\partial_{0}^{2}}D_{a}D^{a}\varrho^{\mu\nu\rho\sigma}\nonumber\\
&&+\Omega_{ij}\left[h_{x}^{i}e_{\mu}^{x}n_{\nu}h_{y}^{j}e_{\rho}^{y}n_{\sigma}\right]D_{a}D^{a}\left[-\left(N^{-1}\bar{\partial_{0}}\right)^{2}\right]\varrho^{\mu\nu\rho\sigma}\nonumber\\
&&+\Omega_{ij}\left[h_{x}^{i}e_{\mu}^{x}n_{\nu}h_{y}^{j}e_{\rho}^{y}n_{\sigma}\right]D_{a}D^{a}D_{b}D^{a}\varrho^{\mu\nu\rho\sigma}\,.
\end{eqnarray}
As a general rule we can write, 
\begin{eqnarray}\label{eq:generalrelationforcommutingoperator1}
&&XDDDDY=D\left(XDDDY\right)-D(X)DDD(Y) \,,\nonumber\\
&&=D\left(D(XDD(Y))-D(X)DD(Y)\right)-D(X)DDD(Y)\,, \nonumber\\
&&=DD(XDD(Y))-D(D(X)DD(Y))-D(X)DDD(Y)\,, \nonumber\\
&&=DD\left(D(XD(Y))-D(X)D(Y)\right)-D(D(X)DD(Y))-D(X)DDD(Y)\,, \nonumber\\
&&=DDD(XD(Y))-DD(D(X)D(Y))-D(D(X)DD(Y))-D(X)DDD(Y)\,, \nonumber\\
&&=DDD\left(D(XY)-D(X)Y\right)-DD(D(X)D(Y))-D(D(X)DD(Y))-D(X)DDD(Y)\,, \nonumber\\
&&=DDDD(XY)-DDD(D(X)Y)-DD(D(X)D(Y))\,, \nonumber\\
&&-D(D(X)DD(Y))-D(X)DDD(Y)\,,
\end{eqnarray} 
where $X$ and $Y$ are some tensors and $D$ is some operator.
Applying this we can write, 
\begin{eqnarray}\label{eq:boundarytermspapereqnwithomegaij}
&&N^{-4}\Omega_{ij}\left[h_{x}^{i}e_{\mu}^{x}n_{\nu}h_{y}^{j}e_{\rho}^{y}n_{\sigma}\right]\bar{\partial_{0}^{4}}\varrho^{\mu\nu\rho\sigma}
\nonumber\\
&&=N^{-4}\Omega_{ij}\{\bar{\partial_{0}^{4}}(\Omega^{ij})-\bar{\partial_{0}^{4}}\left[h_{x}^{i}e_{\mu}^{x}n_{\nu}h_{y}^{j}e_{\rho}^{y}n_{\sigma}\right]\varrho^{\mu\nu\rho\sigma}-\bar{\partial_{0}^{3}}\left[h_{x}^{i}e_{\mu}^{x}n_{\nu}h_{y}^{j}e_{\rho}^{y}n_{\sigma}\right]\bar{\partial_{0}}\varrho^{\mu\nu\rho\sigma}
\nonumber\\
&&-\bar{\partial_{0}^{2}}\left[h_{x}^{i}e_{\mu}^{x}n_{\nu}h_{y}^{j}e_{\rho}^{y}n_{\sigma}\right]\bar{\partial_{0}^{2}}\varrho^{\mu\nu\rho\sigma}-\bar{\partial_{0}}\left[h_{x}^{i}e_{\mu}^{x}n_{\nu}h_{y}^{j}e_{\rho}^{y}n_{\sigma}\right]\bar{\partial_{0}^{3}}\varrho^{\mu\nu\rho\sigma}\}+\cdots
\,,
\end{eqnarray} 
where we dropped the irrelevant terms.
We moreover can generalise the result of (\ref{eq:generalrelationforcommutingoperator1})
and write, 
\begin{eqnarray}
&&X D^{2n} Y= D^{2n}(XY)-D^{2n-1}(D(X)Y)-D^{2n-2}(D(X)D(Y))\,, \nonumber\\
&&-D^{2n-3}(D(X)D^{2}(Y))-\dots-D(D(X)D^{2n-2}(Y))-D(X)D^{2n-1}(Y)
\,.
\end{eqnarray}

\section{Functional Differentiation}\label{C}

Given the constraint equation 
\begin{equation}
2\Psi^{ij}+\frac{\delta f}{\delta \Omega_{ij}}=0,
\end{equation}
suppose that $f=\Omega \mathcal{F}(\Box)\Omega$ and $\mathcal{F}(\Box)=\sum^{\infty}_{n=0}f_{n}\Box^{n}$,
where the coefficients $f_n$ are massless~\footnote{Recall that the $\Box$ term comes
with an associated scale $\Box/M^2$.
}. Then, using the
generalised Euler-Lagrange equations, we have in the coframe (and imposing the condition that $\delta \Omega_{ij}=0$ on the boundary $\partial \mathcal{M}$)
\begin{eqnarray}
\frac{\delta f}{\delta \Omega_{ij}}&+&\frac{\partial
f}{\partial
\Omega_{ij}}-\nabla_{\mu}\left(\frac{\partial
f}{\partial
(\nabla_{\mu}\Omega_{ij})} \right)+\nabla_{\mu}\nabla_{\nu}\left(\frac{\partial
f}{\partial
(\nabla_{\mu}\nabla_{\nu}\Omega_{ij})} \right)+\cdots \nonumber \\
&=&\frac{\partial f}{\partial \Omega_{ij}}+\Box\left(\frac{\partial
f}{\partial (\Box\Omega_{ij})}\right)+\Box^{2}\left(\frac{\partial
f}{\partial (\Box^{2}\Omega_{ij})}\right)+\cdots \nonumber \\
&=&\frac{\partial f}{\partial \Omega_{ij}}+\sum^{\infty}_{n=1}\Box^{n}\left(\frac{\partial
f}{\partial (\Box^{n}\Omega_{ij})}\right) \nonumber \\
&=&f_{0}\frac{\partial(\Omega^{2})}{\partial\Omega_{ij}}+f_{1}\frac{\partial
(\Omega\Box\Omega)}{\partial\Omega_{ij}}+f_{1}\Box\left(\frac{\partial(
\Omega\Box^{}\Omega)}{\partial (\Box\Omega_{ij})}\right)+f_{2}\Box^{2}\left(\frac{\partial
\Omega\Box^{2}\Omega}{\partial (\Box^{2}\Omega_{ij})}\right)+\cdots \nonumber \\
&=&2f_0h^{ij}\Omega+f_{1}h^{ij}\Box\Omega+f_{1}\Box (h^{ij}\Omega)+\cdots \nonumber \\
&=&2f_0h^{ij}\Omega+f_1\Big[ h^{ij}\Box\Omega+\left(\Box \Omega\right) h^{ij}\Big]+\cdots \nonumber \\
&=&2f_0h^{ij}\Omega+2f_1h^{ij}\Box\Omega+\cdots \nonumber \\
&=&2h^{ij}\left( f_0 + f_1 \Box +\cdots\right)\Omega \nonumber \\
&=&2h^{ij}\mathcal{F}(\Box)\Omega \,,
\end{eqnarray}
where we have used that $\Box g^{ij}=\Box h^{ij}=0$.
Note also that:\begin{eqnarray}
f_{1}\Box\left(\frac{\partial(
\Omega\Box^{}\Omega)}{\partial (\Box\Omega_{ij})}\right)=f_{1}\Box \Big(\frac{\partial(
\Omega\Box^{}[h^{mn}\Omega_{mn}])}{\partial (\Box\Omega_{ij})}\Big)\,.
\end{eqnarray}
So we can summarise the results and write, 
\begin{eqnarray}
\frac{\delta (\Omega\Box\Omega)}{\delta \Omega_{ij}}&=&\frac{\partial
(\Omega\Box\Omega)}{\partial\Omega_{ij}}+\Box\left(\frac{\partial(
\Omega\Box^{}\Omega)}{\partial (\Box\Omega_{ij})}\right)\nonumber\\
&=&h^{ij}\Box\Omega+\Box (h^{ij}\Omega)=\Big[ h^{ij}\Box\Omega+\Box\Omega
h^{ij}\Big]=2h^{ij}\Box\Omega\,.
\end{eqnarray}
\begin{eqnarray}
\frac{\delta 
(\Omega_{ij}\Box\Omega^{ij})}{\delta \Omega_{ij}}&=&\frac{\partial
(\Omega_{ij}\Box\Omega^{ij})}{\partial\Omega_{ij}}+\Box\left(\frac{\partial(\Omega_{ij}\Box\Omega^{ij})}{\partial
(\Box\Omega_{ij})}\right)\nonumber\\
&=&\Box\Omega^{ij}+\Box \Omega^{ij}=2\Box\Omega^{ij}\,.
\end{eqnarray}
\begin{eqnarray}
\frac{\delta (\rho\Box\Omega)}{\delta \Omega_{ij}}&=&\Box\left(\frac{\partial(\rho\Box\Omega)}{\partial
(\Box\Omega_{ij})}\right)
=\Box (\rho h^{ij})=h^{ij}\Box\rho\,.
\end{eqnarray}
\begin{eqnarray}
\frac{\delta (\rho_{ij}\Box\Omega^{ij})}{\delta \Omega_{ij}}&=&\Box\left(\frac{\partial(\rho_{ij}\Box\Omega^{ij})}{\partial
(\Box\Omega_{ij})}\right)=\Box \rho^{ij}\,.
\end{eqnarray}
\begin{eqnarray}
\frac{\delta (\Omega\Box\rho)}{\delta \Omega_{ij}}&=&\frac{\partial
(\Omega\Box\rho)}{\partial\Omega_{ij}}=h^{ij}\Box\rho\,.
\end{eqnarray}
\begin{eqnarray}
\frac{\delta (\Omega_{ij}\Box\rho^{ij})}{\delta \Omega_{ij}}&=&\frac{\partial
(\Omega_{ij}\Box\rho^{ij})}{\partial\Omega_{ij}}=\Box \rho^{ij}\,.
\end{eqnarray}
and generalise this to: 
\begin{eqnarray}\label{eq:functionaldifferentationofdifferentterms}
\frac{\delta \big(\Omega\mathcal{F}(\Box)\Omega\big)}{\delta \Omega_{ij}}=2h^{ij}\mathcal{F}(\Box)\Omega,&\quad&
\frac{\delta \big(\Omega_{ij}\mathcal{F}(\Box)\Omega^{ij}\big)}{\delta \Omega_{ij}}=2\mathcal{F}(\Box)\Omega^{ij}\,,\\
\frac{\delta \big(\rho\mathcal{F}(\Box)\Omega\big)}{\delta
\Omega_{ij}}=h^{ij}\mathcal{F}(\Box)\rho,&\quad& \frac{\delta \big(\rho_{ij}\mathcal{F}(\Box)\Omega^{ij}\big)}{\delta
\Omega_{ij}}=\mathcal{F}(\Box)\rho^{ij}\,,\\
\frac{\delta \big(\Omega\mathcal{F}(\Box)\rho\big)}{\delta
\Omega_{ij}}=h^{ij}\mathcal{F}(\Box)\rho,&\quad&
\frac{\delta \big(\Omega_{ij}\mathcal{F}(\Box)\rho^{ij}\big)}{\delta
\Omega_{ij}}=\mathcal{F}(\Box)\rho^{ij}\,.
\end{eqnarray}
\section{Riemann tensor components in ADM gravity}
Using the method of \cite{{Golovnev:2013fj}}, we can find the Riemann tensor components. The Christoffel symbols for the ADM metric in Eq.~(\ref{eq8}) are
\begin{eqnarray}\label{eq:ChristoffelforADM}
       \Gamma_{ij0} &=& \Gamma_{i0j} = - N K_{ij} + D_j \beta_i \nonumber\\
       \Gamma_{ijk} &=& {}^{(3)}\Gamma_{ijk}\nonumber\\
       \Gamma_{00}^0 &=& \frac{1}{N} \left(\dot{N} + \beta^i \partial_i N - \beta^i \beta^j K_{ij} \right)\nonumber\\
        \Gamma^0_{0i} = \Gamma^0_{i0} &=& \frac{1}{N} \left(\partial_i N - \beta^j K_{ij} \right) \nonumber\\
        \Gamma^i_{0j} = \Gamma^i_{j0} &=& - \frac{\beta^i \partial_j N}{N} - N \left( h^{ik} - \frac{\beta^i \beta^k}{N^2} \right) K_{kj} + D_j \beta^i \nonumber\\
        \Gamma^0_{ij} &=& - \frac{1}{N} K_{ij} \nonumber\\
        \Gamma^i_{jk} &=& {}^{(3)} \Gamma^i_{jk} + \frac{\beta^i}{N} K_{jk}
\end{eqnarray}  
where $K_{ij}$ is the extrinsic curvature given by (\ref{eq13}) and in the ADM metric, $N$ is the lapse, $\beta_i$ is the shift and $h_{ij}$ is the induced metric on the hypersurface. Now we can find the Riemann tensor components
\begin{eqnarray}\label{eq:RiemanntensorcompsforADM}
        \mathcal{R}_{ijkl} &=& g_{i\rho} \partial_k \Gamma^\rho_{lj} - g_{i\rho} \partial_l \Gamma^\rho_{kj} + \Gamma_{ik\rho} \Gamma^\rho_{lj} - \Gamma_{il\rho} \Gamma^\rho_{kj} \nonumber\\
        &=& -\beta_i \partial_k \left( \frac{1}{N} K_{jl} \right) + h_{im} \partial_k \left( {}^{(3)} \Gamma^m_{jl} + \frac{\beta^m}{N} K_{jl} \right) - \frac{1}{N} K_{jl} \left( - N K_{ik} + D_k \beta_i \right)\nonumber\\
        && + {}^{(3)} \Gamma_{ikm} \left( {}^{(3)} \Gamma^m_{lj} + \frac{\beta^m}{N} K_{lj} \right) - \left(k \leftrightarrow l \right) \nonumber\\
        &=& R_{ijkl} + K_{ik} K_{jl} - K_{il} K_{jk} \nonumber\\  \end{eqnarray}
 where $R_{ijkl}$ is the Riemann tensor of the induced metric on the hypersurface. Then
 \begin{eqnarray}
        n_\mu {\mathcal{R}^\mu}_{ijk} &=& - N \left( \partial_j \Gamma^0_{ki} + \Gamma^0_{j\rho}\Gamma^\rho_{ki}\right) - \left( j \leftrightarrow k \right) \nonumber\\
        &=& \partial_j K_{ki} + {}^{(3)} \Gamma^m_{ki} K_{jm} - \left( j \leftrightarrow k \right) \nonumber\\
        &=& D_j K_{ki} - D_k K_{ji} 
\end{eqnarray}
Relabelling the indices, we obtain that
\begin{equation}
n^{\mu}\mathcal{R}_{ijk \mu}= D_{j}K_{ki}-D_{i}K_{jk}
\end{equation}
Finally, we have that 
\begin{eqnarray}
        n_\mu {\mathcal{R}^\mu}_{i0j} &=& n_\mu \left( \partial_0 \Gamma^\mu_{ji} - \partial_j \Gamma^\mu_{0i} + \Gamma^\mu_{0\rho} \Gamma^\rho_{ji} - \Gamma^\mu_{j\rho} \Gamma^\rho_{0i} \right) \nonumber\\
        &=& \dot{K}_{ij} + D_i D_j N + N {K_i}^k K_{kj} - D_j \left( K_{ik} \beta^k \right) - K_{kj} D_i \beta^k        
\end{eqnarray}

Hence
\begin{eqnarray} 
        n^\mu n^\nu \mathcal{R}_{\mu i \nu j} &=& n^0 n^\mu \mathcal{R}_{\mu i0 j}+ n^k n^\mu \mathcal{R}_{\mu i kj } \nonumber\\
        &=& \frac{1}{N} \left( \dot{K}_{ij} + D_i D_j N + N {K_i}^k K_{kj}- D_j \left( K_{ik} \beta^k \right) - K_{kj} D_i \beta^k \right)\nonumber\\
        && + \frac{\beta^k}{N^2} \left(D_j K_{ki} - D_k K_{ji}\right)\nonumber\\
        &=&\frac{1}{N} \left( \dot{K}_{ij} + D_i D_j N + N {K_i}^k K_{kj} - \pounds_\beta K_{ij} \right)    
\end{eqnarray}
where $\pounds_\beta K_{ij}\equiv \beta^{k}D_k K_{ij}+K_{ik}D_j\beta^k+K_{jk}D_i\beta^k$.\ Therefore overall, we have 
\begin{eqnarray}
\mathcal{R}_{ijkl}&\equiv&K_{ik}K_{jl}-K_{il}K_{jk}+R_{ijkl}\,, \\ \label{xx5}
\mathcal{R}_{ijk\mathbf{n}}&\equiv&n^{\mu}\mathcal{R}_{ijk\mu}=D_{j}K_{ik}-D_{i}K_{jk}\,,
\\ \label{xx2}
\mathcal{R}_{i\mathbf{n}j\mathbf{n}}&\equiv&n^{\mu}n^{\nu}\mathcal{R}_{i\mu
j\nu}=N^{-1}\big(\partial_{t}K_{ij}-\mathsterling_\beta
K_{ij}\big)+K_{ik}K^{\ k}_{j}+N^{-1}D_i D_jN\,,
\end{eqnarray}
 
\subsection{Coframe}

Since in the coframe slicing Eq.~(\ref{eq16}) we have $g^{0i}=g_{0i}=0$, therefore from Eq.~(\ref{eq9}), we also
have $n^i=n_i=0$ ($n_{0}$ and $n^0$ stay the same
as in Eq.~\eqref{eq9}). Then the Christoffel symbols become~\footnote{In the coframe slicing when we write $\partial_\mu$ we mean that $\partial_\mu$ is $\bar\partial_0$ when $\mu=0$ and $\partial_\mu$
is $\partial_i$ when $\mu=i$. }
\begin{eqnarray}
        \Gamma^0_{00} &=& \frac{1}{2} g^{0\mu} \left(\bar \partial_0 g_{\mu 0}
+ \bar \partial_0 g_{0 \mu} - \partial_\mu g_{00} \right) \nonumber \\
        &=& \frac{1}{2} g^{00}  \partial_0 g_{0 0} 
        = \frac{1}{2} \left( - \frac{1}{N^2} \right) \partial_0 \left(
- N^2 \right) \nonumber \\
        &=& \frac{\partial_0 N}{N}, 
\end{eqnarray}
\begin{eqnarray}
        \Gamma^0_{0i} = \Gamma^0_{i0} &=& \frac{1}{2} g^{0\mu} \left(\bar \partial_0
g_{\mu i} + \partial_i g_{\mu 0} - \partial_\mu g_{i0} \right) \nonumber
\\
        &=& \frac{1}{2} g^{00} \partial_i g_{00} 
        = \frac{1}{2} \left( \frac{-1}{N^2} \right) \partial_i \left( - N^2  \right) \\
\nonumber
        &=& \frac{\partial_i N }{N}, 
\end{eqnarray}
\begin{eqnarray}
        \Gamma^i_{00} &=& \frac{1}{2} g^{i\mu} \left(\bar \partial_0 g_{\mu 0}
+\bar \partial_0 g_{\mu 0} - \partial_\mu g_{00} \right) \nonumber \\
        &=& - \frac{1}{2} g^{ij} \partial_j g_{00} 
        = - \frac{1}{2} h^{ij} \partial_j \left( - N^2 \right) \nonumber\\
        &=&  N h^{ij} \partial_j N,
\end{eqnarray}
\begin{eqnarray}
        \Gamma^i_{j0} &=& \frac{1}{2} g^{i \mu} \left(\bar \partial_0 g_{\mu j} 
        + \partial_j g_{\mu 0} - \partial_\mu g_{j 0} \right) \\  \nonumber
        &=& \frac{1}{2} g^{ik} \left(\bar \partial_0 g_{kj} \right) \nonumber\\
        &=& \frac{1}{2} h^{ik}\bar \partial_0 h_{jk}, 
\end{eqnarray}
\begin{eqnarray}
        \Gamma^0_{ij} &=& \frac{1}{2} g^{0\mu} \left( \partial_j g_{\mu i}
+ \partial_i g_{\mu j} - \partial_\mu g_{ij} \right) \nonumber \\
        &=& - \frac{1}{2} g^{00}\bar \partial_0 g_{ij} 
        = - \frac{1}{2}\left( \frac{-1}{N^2} \right)\bar \partial_0 h_{ij} \nonumber \\
        &=& \frac{1}{2} \frac{1}{N^2}\bar \partial_0 h_{ij},
\end{eqnarray}
\begin{eqnarray}
        \Gamma^i_{jk} &=& \frac{1}{2} g^{i\mu} \left( \partial_j g_{\mu k}
+ \partial_k g_{\mu j} - \partial_\mu g_{kj} \right) \nonumber \\
        &=& \frac{1}{2} h^{il} \left( \partial_j h_{lk} + \partial_k h_{lj}
- \partial_l h_{jk} \right).
\end{eqnarray}
To summarise
\begin{eqnarray}
        \Gamma^0_{00} &=& \frac{\partial_0 N}{N}, \quad \quad \quad \quad
        \Gamma^0_{0i} =  \frac{\partial_i N}{N}, \nonumber \\
        \Gamma^i_{00} &=&  N h^{ij} \partial_j N, \quad \quad \quad
        \Gamma^i_{j0} = \frac{1}{2} h^{ik}\bar \partial_0 h_{jk}, \nonumber\\
        \Gamma^0_{ij} &=& \frac{1}{2} \frac{1}{N^2}\bar \partial_0 h_{ij}, \quad \quad
        \Gamma^i_{jk} =  \frac{1}{2} h^{il} \left( \partial_j h_{lk} +
\partial_k h_{lj} - \partial_l h_{jk} \right).
\end{eqnarray}
Then using Eq.~(\ref{eq22}) and Eq.~(\ref{eq23}), we can find the $\gamma^\mu_{\nu\rho}$s, the analogues of the Christoffel symbols in the coframe.
\begin{eqnarray}
        \gamma^{i}_{jk} = \Gamma^i_{jk}, 
        \quad \gamma^i_{0k} = - N {K^i}_k,
        \quad \gamma^{i}_{j0} = - N {K^i}_k + \partial_j \beta^i,
        \quad \gamma^{0}_{ij} = - N^{-1} K_{ij},\nonumber \\
        \gamma^{i}_{00} = N \partial^i N, 
        \quad \gamma^{0}_{0i} = \gamma^{0}_{i0} = \partial_i \log N,
        \quad \gamma^{0}_{00} = \partial_0 \log N \quad
\end{eqnarray}
Then using the same method as in Eq.~(\ref{eq:RiemanntensorcompsforADM}),
\begin{eqnarray}
        \mathcal{R}_{ijkl} &=& g_{i\rho} \partial_k \gamma^{\rho}_{lj} - g_{i\rho} \partial_l \gamma^{\rho}_{kj} + \gamma_{ik\rho} 
        \gamma^\rho_{lj} -  \gamma_{il\rho} \gamma^{\rho}_{kj}\nonumber\\
        &=& R_{ijkl} + K_{ik} K_{jl} - K_{il} K_{jk} 
\end{eqnarray}
 Next
 \begin{eqnarray}
        \mathcal{R}_{0ijk} &=& - N^{2} \left( \partial_j \gamma^0_{ki} + \gamma^0_{j\rho}\gamma^\rho_{ki}\right)
- \left( j \leftrightarrow k \right) \nonumber\\
       &=& N \left( D_j K_{ki} - D_k K_{ji} \right)
\end{eqnarray}
Finally we have that in the coframe,
\begin{eqnarray} 
         \mathcal{R}_{0 i 0 j} &=&-N^{2} \left( \bar \partial_0 \gamma_{ji}^{0}
-\partial_j \gamma_{0i}^{0} + \gamma_{0\rho}^{0} \gamma^\rho_{ji} - \gamma_{j\rho}^{0}
\gamma^\rho_{0i} \right) \nonumber\\
        &=& N \left( \bar \partial_{0}K_{ij}  + N {K_i}^k K_{kj}+ D_i D_j N\right) 
\end{eqnarray}
Hence the non-vanishing components of the Riemann
tensor in the coframe, namely the Gauss, Codazzi and Ricci tensor, become: \begin{eqnarray}
\mathcal{R}_{ijkl}&=&K_{ik}K_{jl}-K_{il}K_{jk}+R_{ijkl}\,,\nonumber \\
\mathcal{R}_{0ijk}&=&N(D_{j}K_{ki}-D_{k}K_{ji})\,, \nonumber
\\
\mathcal{R}_{0i0j}&=&N(\bar \partial_{0} K_{ij}+NK_{ik}K^{\ k}_{j}+D_i D_jN\,),
\end{eqnarray}    
where $K_{ij}$ is the extrinsic curvature of the hypersurface, given in the coframe by Eq.~(\ref{eq29}) and $R_{ijkl}$ is the Riemann tensor of the induced metric on the hypersurface.



\end{document}